\documentclass{article}

\usepackage{subfig}
\usepackage{array}
\usepackage{longtable}
\usepackage[inline]{enumitem}
\usepackage[hidelinks]{hyperref}

\usepackage{tikz}
\usetikzlibrary{shapes.geometric, arrows, calc}
\usepackage{amsmath}
\usepackage[left=2cm,right=2cm]{geometry}

\newcommand{\gulyas}{Andr\'as~Guly\'as}
\newcommand{\nemeth}{Felici\'an~N\'emeth}
\newcommand{\pelle}{Istv\'an~Pelle}

\newcommand{\epoxide}{Epoxide}
\newcommand{\emacs}{Emacs}

\newcommand{\tsgFile}{\texttt{.tsg}}
\newcommand{\scope}{\texttt{::}}
\newcommand{\link}{\texttt{->}}
\newcommand{\viewLink}{\texttt{-->}}
\newcommand{\eldoc}{ElDoc}
\newcommand{\autocomplete}{Auto Complete}
\newcommand{\ido}{Ido}
\newcommand{\ibuffer}{Ibuffer}
\newcommand{\cogre}{COGRE}
\newcommand{\graphviz}{Graphviz}

\newcommand{\iptables}{\texttt{iptables}}
\newcommand{\ssh}{SSH}
\newcommand{\texttsg}[1]{\texttt{#1}}
\newcommand{\emacslisp}{\emacs{} Lisp}

\usepackage[acronym,nomain]{glossaries}
\setacronymstyle{long-short}

\usepackage[super]{nth}
\usepackage{listings}

\usepackage{authblk}

\makeatletter

\pgfdeclareshape{decision}{
  \savedanchor\northeast{%
    \pgfmathsetlength\pgf@x{\pgfshapeminwidth}%
    \pgfmathsetlength\pgf@y{\pgfshapeminheight}%
    \pgf@x=0.5\pgf@x
    \pgf@y=0.5\pgf@y
  }
  \savedanchor\southwest{%
    \pgfmathsetlength\pgf@x{\pgfshapeminwidth}%
    \pgfmathsetlength\pgf@y{\pgfshapeminheight}%
    \pgf@x=-0.5\pgf@x
    \pgf@y=-0.5\pgf@y
  }
  \inheritanchorborder[from=rectangle]

  \anchor{center}{\pgfpointorigin}
  \anchor{north}{\northeast \pgf@x=0pt}
  \anchor{east}{\northeast \pgf@y=0pt}
  \anchor{south}{\southwest \pgf@x=0pt}
  \anchor{west}{\southwest \pgf@y=0pt}
  \anchor{north east}{\northeast}
  \anchor{north west}{\northeast \pgf@x=-\pgf@x}
  \anchor{south west}{\southwest}
  \anchor{south east}{\southwest \pgf@x=-\pgf@x}
  \anchor{text}{
    \pgfpointorigin
    \advance\pgf@x by -.5\wd\pgfnodeparttextbox%
    \advance\pgf@y by -.5\ht\pgfnodeparttextbox%
    \advance\pgf@y by +.5\dp\pgfnodeparttextbox%
  }

  \anchor{inputa}{
    \pgf@process{\northeast}%
    \pgf@x=-1\pgf@x%
    \pgf@y=.5\pgf@y%
  }
  \anchor{outputa}{
    \pgf@process{\northeast}%
    \pgf@y=.5\pgf@y%
  }
  \anchor{outputb}{
    \pgf@process{\northeast}%
    \pgf@y=0\pgf@y%
  }
  \anchor{outputc}{
    \pgf@process{\northeast}%
    \pgf@y=-.5\pgf@y%
  }
  \backgroundpath{
    \pgfpathrectanglecorners{\southwest}{\northeast}

    \begingroup
    \tikzset{decision node/port labels} % Use font from this style
    \tikz@textfont

    \pgf@anchor@decision@inputa
    \pgftext[left,base,at={\pgfpoint{\pgf@x}{\pgf@y}},x=\pgfshapeinnerxsep]{\raisebox{-0.75ex}{I0}}

    \pgf@anchor@decision@outputa
    \pgftext[right,base,at={\pgfpoint{\pgf@x}{\pgf@y}},x=-\pgfshapeinnerxsep]{\raisebox{-.75ex}{O0}}

    \pgf@anchor@decision@outputb
    \pgftext[right,base,at={\pgfpoint{\pgf@x}{\pgf@y}},x=-\pgfshapeinnerxsep]{\raisebox{-.75ex}{O1}}

    \pgf@anchor@decision@outputc
    \pgftext[right,base,at={\pgfpoint{\pgf@x}{\pgf@y}},x=-\pgfshapeinnerxsep]{\raisebox{-.75ex}{O2}}

    \endgroup
  }
}

\pgfdeclareshape{nodea}{
  \savedanchor\northeast{%
    \pgfmathsetlength\pgf@x{\pgfshapeminwidth}%
    \pgfmathsetlength\pgf@y{\pgfshapeminheight}%
    \pgf@x=0.5\pgf@x
    \pgf@y=0.5\pgf@y
  }
  \savedanchor\southwest{%
    \pgfmathsetlength\pgf@x{\pgfshapeminwidth}%
    \pgfmathsetlength\pgf@y{\pgfshapeminheight}%
    \pgf@x=-0.5\pgf@x
    \pgf@y=-0.5\pgf@y
  }
  \inheritanchorborder[from=rectangle]

  \anchor{center}{\pgfpointorigin}
  \anchor{north}{\northeast \pgf@x=0pt}
  \anchor{east}{\northeast \pgf@y=0pt}
  \anchor{south}{\southwest \pgf@x=0pt}
  \anchor{west}{\southwest \pgf@y=0pt}
  \anchor{north east}{\northeast}
  \anchor{north west}{\northeast \pgf@x=-\pgf@x}
  \anchor{south west}{\southwest}
  \anchor{south east}{\southwest \pgf@x=-\pgf@x}
  \anchor{text}{
    \pgfpointorigin
    \advance\pgf@x by -.5\wd\pgfnodeparttextbox%
    \advance\pgf@y by -.5\ht\pgfnodeparttextbox%
    \advance\pgf@y by +.5\dp\pgfnodeparttextbox%
  }

  \anchor{outputa}{
    \pgf@process{\northeast}%
    \pgf@y=.5\pgf@y%
  }
  \backgroundpath{
    \pgfpathrectanglecorners{\southwest}{\northeast}

    \begingroup
    \tikzset{nodea node/port labels} % Use font from this style
    \tikz@textfont

    \pgf@anchor@nodea@outputa
    \pgftext[right,base,at={\pgfpoint{\pgf@x}{\pgf@y}},x=-\pgfshapeinnerxsep]{\raisebox{-.75ex}{O0}}

    \endgroup
  }
}

\pgfdeclareshape{nodeb}{
  \savedanchor\northeast{%
    \pgfmathsetlength\pgf@x{\pgfshapeminwidth}%
    \pgfmathsetlength\pgf@y{\pgfshapeminheight}%
    \pgf@x=0.5\pgf@x
    \pgf@y=0.5\pgf@y
  }
  \savedanchor\southwest{%
    \pgfmathsetlength\pgf@x{\pgfshapeminwidth}%
    \pgfmathsetlength\pgf@y{\pgfshapeminheight}%
    \pgf@x=-0.5\pgf@x
    \pgf@y=-0.5\pgf@y
  }
  \inheritanchorborder[from=rectangle]

  \anchor{center}{\pgfpointorigin}
  \anchor{north}{\northeast \pgf@x=0pt}
  \anchor{east}{\northeast \pgf@y=0pt}
  \anchor{south}{\southwest \pgf@x=0pt}
  \anchor{west}{\southwest \pgf@y=0pt}
  \anchor{north east}{\northeast}
  \anchor{north west}{\northeast \pgf@x=-\pgf@x}
  \anchor{south west}{\southwest}
  \anchor{south east}{\southwest \pgf@x=-\pgf@x}
  \anchor{text}{
    \pgfpointorigin
    \advance\pgf@x by -.5\wd\pgfnodeparttextbox%
    \advance\pgf@y by -.5\ht\pgfnodeparttextbox%
    \advance\pgf@y by +.5\dp\pgfnodeparttextbox%
  }

  \anchor{inputa}{
    \pgf@process{\northeast}%
    \pgf@x=-1\pgf@x%
    \pgf@y=.5\pgf@y%
  }
  \anchor{outputa}{
    \pgf@process{\northeast}%
    \pgf@y=.5\pgf@y%
  }
  \backgroundpath{
    \pgfpathrectanglecorners{\southwest}{\northeast}

    \begingroup
    \tikzset{nodeb node/port labels} % Use font from this style
    \tikz@textfont

    \pgf@anchor@nodeb@inputa
    \pgftext[left,base,at={\pgfpoint{\pgf@x}{\pgf@y}},x=\pgfshapeinnerxsep]{\raisebox{-0.75ex}{I0}}

    \pgf@anchor@nodeb@outputa
    \pgftext[right,base,at={\pgfpoint{\pgf@x}{\pgf@y}},x=-\pgfshapeinnerxsep]{\raisebox{-.75ex}{O0}}

    \endgroup
  }
}

\pgfdeclareshape{nodec}{
  \savedanchor\northeast{%
    \pgfmathsetlength\pgf@x{\pgfshapeminwidth}%
    \pgfmathsetlength\pgf@y{\pgfshapeminheight}%
    \pgf@x=0.5\pgf@x
    \pgf@y=0.5\pgf@y
  }
  \savedanchor\southwest{%
    \pgfmathsetlength\pgf@x{\pgfshapeminwidth}%
    \pgfmathsetlength\pgf@y{\pgfshapeminheight}%
    \pgf@x=-0.5\pgf@x
    \pgf@y=-0.5\pgf@y
  }
  \inheritanchorborder[from=rectangle]

  \anchor{center}{\pgfpointorigin}
  \anchor{north}{\northeast \pgf@x=0pt}
  \anchor{east}{\northeast \pgf@y=0pt}
  \anchor{south}{\southwest \pgf@x=0pt}
  \anchor{west}{\southwest \pgf@y=0pt}
  \anchor{north east}{\northeast}
  \anchor{north west}{\northeast \pgf@x=-\pgf@x}
  \anchor{south west}{\southwest}
  \anchor{south east}{\southwest \pgf@x=-\pgf@x}
  \anchor{text}{
    \pgfpointorigin
    \advance\pgf@x by -.5\wd\pgfnodeparttextbox%
    \advance\pgf@y by -.5\ht\pgfnodeparttextbox%
    \advance\pgf@y by +.5\dp\pgfnodeparttextbox%
  }

  \anchor{inputa}{
    \pgf@process{\northeast}%
    \pgf@x=-1\pgf@x%
    \pgf@y=.5\pgf@y%
  }
  \anchor{outputa}{
    \pgf@process{\northeast}%
    \pgf@y=.5\pgf@y%
  }
  \anchor{confc}{
    \pgf@process{\northeast}%
    \pgf@x=.5\pgf@x%
  }
  \backgroundpath{
    \pgfpathrectanglecorners{\southwest}{\northeast}

    \begingroup
    \tikzset{nodec node/port labels} % Use font from this style
    \tikz@textfont

    \pgf@anchor@nodec@inputa
    \pgftext[left,base,at={\pgfpoint{\pgf@x}{\pgf@y}},x=\pgfshapeinnerxsep]{\raisebox{-0.75ex}{I0}}

    \pgf@anchor@nodec@outputa
    \pgftext[right,base,at={\pgfpoint{\pgf@x}{\pgf@y}},x=-\pgfshapeinnerxsep]{\raisebox{-.75ex}{O0}}

    \pgf@anchor@nodec@confc
    \pgftext[top,at={\pgfpoint{\pgf@x}{\pgf@y}},y=-\pgfshapeinnerysep]{C3}

    \endgroup
  }
}

\pgfdeclareshape{noded}{
  \savedanchor\northeast{%
    \pgfmathsetlength\pgf@x{\pgfshapeminwidth}%
    \pgfmathsetlength\pgf@y{\pgfshapeminheight}%
    \pgf@x=0.5\pgf@x
    \pgf@y=0.5\pgf@y
  }
  \savedanchor\southwest{%
    \pgfmathsetlength\pgf@x{\pgfshapeminwidth}%
    \pgfmathsetlength\pgf@y{\pgfshapeminheight}%
    \pgf@x=-0.5\pgf@x
    \pgf@y=-0.5\pgf@y
  }
  \inheritanchorborder[from=rectangle]

  \anchor{center}{\pgfpointorigin}
  \anchor{north}{\northeast \pgf@x=0pt}
  \anchor{east}{\northeast \pgf@y=0pt}
  \anchor{south}{\southwest \pgf@x=0pt}
  \anchor{west}{\southwest \pgf@y=0pt}
  \anchor{north east}{\northeast}
  \anchor{north west}{\northeast \pgf@x=-\pgf@x}
  \anchor{south west}{\southwest}
  \anchor{south east}{\southwest \pgf@x=-\pgf@x}
  \anchor{text}{
    \pgfpointorigin
    \advance\pgf@x by -.5\wd\pgfnodeparttextbox%
    \advance\pgf@y by -.5\ht\pgfnodeparttextbox%
    \advance\pgf@y by +.5\dp\pgfnodeparttextbox%
  }

  \anchor{inputa}{
    \pgf@process{\northeast}%
    \pgf@x=-1\pgf@x%
    \pgf@y=.5\pgf@y%
  }
  \anchor{outputa}{
    \pgf@process{\northeast}%
    \pgf@y=.5\pgf@y%
  }
  \anchor{confa}{
    \pgf@process{\northeast}%
    \pgf@x=-.5\pgf@x%
  }
  \backgroundpath{
    \pgfpathrectanglecorners{\southwest}{\northeast}

    \begingroup
    \tikzset{noded node/port labels} % Use font from this style
    \tikz@textfont

    \pgf@anchor@noded@inputa
    \pgftext[left,base,at={\pgfpoint{\pgf@x}{\pgf@y}},x=\pgfshapeinnerxsep]{\raisebox{-0.75ex}{I0}}

    \pgf@anchor@noded@outputa
    \pgftext[right,base,at={\pgfpoint{\pgf@x}{\pgf@y}},x=-\pgfshapeinnerxsep]{\raisebox{-.75ex}{O0}}

    \pgf@anchor@noded@confa
    \pgftext[top,at={\pgfpoint{\pgf@x}{\pgf@y}},y=-\pgfshapeinnerysep]{C1}

    \endgroup
  }
}

\pgfdeclareshape{nodee}{
  \savedanchor\northeast{%
    \pgfmathsetlength\pgf@x{\pgfshapeminwidth}%
    \pgfmathsetlength\pgf@y{\pgfshapeminheight}%
    \pgf@x=0.5\pgf@x
    \pgf@y=0.5\pgf@y
  }
  \savedanchor\southwest{%
    \pgfmathsetlength\pgf@x{\pgfshapeminwidth}%
    \pgfmathsetlength\pgf@y{\pgfshapeminheight}%
    \pgf@x=-0.5\pgf@x
    \pgf@y=-0.5\pgf@y
  }
  \inheritanchorborder[from=rectangle]

  \anchor{center}{\pgfpointorigin}
  \anchor{north}{\northeast \pgf@x=0pt}
  \anchor{east}{\northeast \pgf@y=0pt}
  \anchor{south}{\southwest \pgf@x=0pt}
  \anchor{west}{\southwest \pgf@y=0pt}
  \anchor{north east}{\northeast}
  \anchor{north west}{\northeast \pgf@x=-\pgf@x}
  \anchor{south west}{\southwest}
  \anchor{south east}{\southwest \pgf@x=-\pgf@x}
  \anchor{text}{
    \pgfpointorigin
    \advance\pgf@x by -.5\wd\pgfnodeparttextbox%
    \advance\pgf@y by -.5\ht\pgfnodeparttextbox%
    \advance\pgf@y by +.5\dp\pgfnodeparttextbox%
  }

  \anchor{inputa}{
    \pgf@process{\northeast}%
    \pgf@x=-1\pgf@x%
    \pgf@y=.5\pgf@y%
  }
  \anchor{outputa}{
    \pgf@process{\northeast}%
    \pgf@y=.5\pgf@y%
  }
  \anchor{outputb}{
    \pgf@process{\northeast}%
    \pgf@y=0\pgf@y%
  }
  \backgroundpath{
    \pgfpathrectanglecorners{\southwest}{\northeast}

    \begingroup
    \tikzset{nodee node/port labels} % Use font from this style
    \tikz@textfont

    \pgf@anchor@nodee@inputa
    \pgftext[left,base,at={\pgfpoint{\pgf@x}{\pgf@y}},x=\pgfshapeinnerxsep]{\raisebox{-0.75ex}{I0}}

    \pgf@anchor@nodee@outputa
    \pgftext[right,base,at={\pgfpoint{\pgf@x}{\pgf@y}},x=-\pgfshapeinnerxsep]{\raisebox{-.75ex}{O0}}
    
    \pgf@anchor@nodee@outputb
    \pgftext[right,base,at={\pgfpoint{\pgf@x}{\pgf@y}},x=-\pgfshapeinnerxsep]{\raisebox{-.75ex}{O1}}

    \endgroup
  }
}

\pgfdeclareshape{nodef}{
  \savedanchor\northeast{%
    \pgfmathsetlength\pgf@x{\pgfshapeminwidth}%
    \pgfmathsetlength\pgf@y{\pgfshapeminheight}%
    \pgf@x=0.5\pgf@x
    \pgf@y=0.5\pgf@y
  }
  \savedanchor\southwest{%
    \pgfmathsetlength\pgf@x{\pgfshapeminwidth}%
    \pgfmathsetlength\pgf@y{\pgfshapeminheight}%
    \pgf@x=-0.5\pgf@x
    \pgf@y=-0.5\pgf@y
  }
  \inheritanchorborder[from=rectangle]

  \anchor{center}{\pgfpointorigin}
  \anchor{north}{\northeast \pgf@x=0pt}
  \anchor{east}{\northeast \pgf@y=0pt}
  \anchor{south}{\southwest \pgf@x=0pt}
  \anchor{west}{\southwest \pgf@y=0pt}
  \anchor{north east}{\northeast}
  \anchor{north west}{\northeast \pgf@x=-\pgf@x}
  \anchor{south west}{\southwest}
  \anchor{south east}{\southwest \pgf@x=-\pgf@x}
  \anchor{text}{
    \pgfpointorigin
    \advance\pgf@x by -.5\wd\pgfnodeparttextbox%
    \advance\pgf@y by -.5\ht\pgfnodeparttextbox%
    \advance\pgf@y by +.5\dp\pgfnodeparttextbox%
  }

  \anchor{inputa}{
    \pgf@process{\northeast}%
    \pgf@x=-1\pgf@x%
    \pgf@y=.5\pgf@y%
  }
  \backgroundpath{
    \pgfpathrectanglecorners{\southwest}{\northeast}

    \begingroup
    \tikzset{nodef node/port labels} % Use font from this style
    \tikz@textfont

    \pgf@anchor@nodef@inputa
    \pgftext[left,base,at={\pgfpoint{\pgf@x}{\pgf@y}},x=\pgfshapeinnerxsep]{\raisebox{-0.75ex}{I0}}

    \endgroup
  }
}

\pgfdeclareshape{nodeg}{
  \savedanchor\northeast{%
    \pgfmathsetlength\pgf@x{\pgfshapeminwidth}%
    \pgfmathsetlength\pgf@y{\pgfshapeminheight}%
    \pgf@x=0.5\pgf@x
    \pgf@y=0.5\pgf@y
  }
  \savedanchor\southwest{%
    \pgfmathsetlength\pgf@x{\pgfshapeminwidth}%
    \pgfmathsetlength\pgf@y{\pgfshapeminheight}%
    \pgf@x=-0.5\pgf@x
    \pgf@y=-0.5\pgf@y
  }
  \inheritanchorborder[from=rectangle]

  \anchor{center}{\pgfpointorigin}
  \anchor{north}{\northeast \pgf@x=0pt}
  \anchor{east}{\northeast \pgf@y=0pt}
  \anchor{south}{\southwest \pgf@x=0pt}
  \anchor{west}{\southwest \pgf@y=0pt}
  \anchor{north east}{\northeast}
  \anchor{north west}{\northeast \pgf@x=-\pgf@x}
  \anchor{south west}{\southwest}
  \anchor{south east}{\southwest \pgf@x=-\pgf@x}
  \anchor{text}{
    \pgfpointorigin
    \advance\pgf@x by -.5\wd\pgfnodeparttextbox%
    \advance\pgf@y by -.5\ht\pgfnodeparttextbox%
    \advance\pgf@y by +.5\dp\pgfnodeparttextbox%
  }

  \anchor{inputa}{
    \pgf@process{\northeast}%
    \pgf@x=-1\pgf@x%
    \pgf@y=.5\pgf@y%
  }
  \anchor{inputb}{
    \pgf@process{\northeast}%
    \pgf@x=-1\pgf@x%
    \pgf@y=0\pgf@y%
  }
  \anchor{outputa}{
    \pgf@process{\northeast}%
    \pgf@y=.5\pgf@y%
  }
  \backgroundpath{
    \pgfpathrectanglecorners{\southwest}{\northeast}

    \begingroup
    \tikzset{nodeg node/port labels} % Use font from this style
    \tikz@textfont

    \pgf@anchor@nodeg@inputa
    \pgftext[left,base,at={\pgfpoint{\pgf@x}{\pgf@y}},x=\pgfshapeinnerxsep]{\raisebox{-0.75ex}{I0}}

    \pgf@anchor@nodeg@inputb
    \pgftext[left,base,at={\pgfpoint{\pgf@x}{\pgf@y}},x=\pgfshapeinnerxsep]{\raisebox{-0.75ex}{I1}}

    \pgf@anchor@nodeg@outputa
    \pgftext[right,base,at={\pgfpoint{\pgf@x}{\pgf@y}},x=-\pgfshapeinnerxsep]{\raisebox{-.75ex}{O0}}

    \endgroup
  }
}

\pgfdeclareshape{summary}{
  \savedanchor\northeast{%
    \pgfmathsetlength\pgf@x{\pgfshapeminwidth}%
    \pgfmathsetlength\pgf@y{\pgfshapeminheight}%
    \pgf@x=0.5\pgf@x
    \pgf@y=0.5\pgf@y
  }
  \savedanchor\southwest{%
    \pgfmathsetlength\pgf@x{\pgfshapeminwidth}%
    \pgfmathsetlength\pgf@y{\pgfshapeminheight}%
    \pgf@x=-0.5\pgf@x
    \pgf@y=-0.5\pgf@y
  }
  \inheritanchorborder[from=rectangle]

  \anchor{center}{\pgfpointorigin}
  \anchor{north}{\northeast \pgf@x=0pt}
  \anchor{east}{\northeast \pgf@y=0pt}
  \anchor{south}{\southwest \pgf@x=0pt}
  \anchor{west}{\southwest \pgf@y=0pt}
  \anchor{north east}{\northeast}
  \anchor{north west}{\northeast \pgf@x=-\pgf@x}
  \anchor{south west}{\southwest}
  \anchor{south east}{\southwest \pgf@x=-\pgf@x}
  \anchor{text}{
    \pgfpointorigin
    \advance\pgf@x by -.5\wd\pgfnodeparttextbox%
    \advance\pgf@y by -.5\ht\pgfnodeparttextbox%
    \advance\pgf@y by +.5\dp\pgfnodeparttextbox%
  }

  \anchor{inputa}{
    \pgf@process{\northeast}%
    \pgf@x=-1\pgf@x%
    \pgf@y=.8\pgf@y%
  }
  \anchor{inputb}{
    \pgf@process{\northeast}%
    \pgf@x=-1\pgf@x%
    \pgf@y=.5\pgf@y%
  }
  \anchor{inputc}{
    \pgf@process{\northeast}%
    \pgf@x=-1\pgf@x%
    \pgf@y=.2\pgf@y%
  }
  \anchor{inputd}{
    \pgf@process{\northeast}%
    \pgf@x=-1\pgf@x%
    \pgf@y=-0.1\pgf@y%
  }
  \anchor{inpute}{
    \pgf@process{\northeast}%
    \pgf@x=-1\pgf@x%
    \pgf@y=-.4\pgf@y%
  }
  \anchor{inputf}{
    \pgf@process{\northeast}%
    \pgf@x=-1\pgf@x%
    \pgf@y=-.7\pgf@y%
  }
  \backgroundpath{
    \pgfpathrectanglecorners{\southwest}{\northeast}

    \begingroup
    \tikzset{summary node/port labels} % Use font from this style
    \tikz@textfont

    \pgf@anchor@summary@inputa
    \pgftext[left,base,at={\pgfpoint{\pgf@x}{\pgf@y}},x=\pgfshapeinnerxsep]{\raisebox{-0.75ex}{I0}}
    
    \pgf@anchor@summary@inputb
    \pgftext[left,base,at={\pgfpoint{\pgf@x}{\pgf@y}},x=\pgfshapeinnerxsep]{\raisebox{-0.75ex}{I1}}

    \pgf@anchor@summary@inputc
    \pgftext[left,base,at={\pgfpoint{\pgf@x}{\pgf@y}},x=\pgfshapeinnerxsep]{\raisebox{-0.75ex}{I2}}

    \pgf@anchor@summary@inputd
    \pgftext[left,base,at={\pgfpoint{\pgf@x}{\pgf@y}},x=\pgfshapeinnerxsep]{\raisebox{-0.75ex}{I3}}

    \pgf@anchor@summary@inpute
    \pgftext[left,base,at={\pgfpoint{\pgf@x}{\pgf@y}},x=\pgfshapeinnerxsep]{\raisebox{-0.75ex}{I4}}

    \pgf@anchor@summary@inputf
    \pgftext[left,base,at={\pgfpoint{\pgf@x}{\pgf@y}},x=\pgfshapeinnerxsep]{\raisebox{-0.75ex}{I5}}

    \endgroup
  }
}

\pgfdeclareshape{table}{
  \savedanchor\northeast{%
    \pgfmathsetlength\pgf@x{\pgfshapeminwidth}%
    \pgfmathsetlength\pgf@y{\pgfshapeminheight}%
    \pgf@x=0.5\pgf@x
    \pgf@y=0.5\pgf@y
  }
  \savedanchor\southwest{%
    \pgfmathsetlength\pgf@x{\pgfshapeminwidth}%
    \pgfmathsetlength\pgf@y{\pgfshapeminheight}%
    \pgf@x=-0.5\pgf@x
    \pgf@y=-0.5\pgf@y
  }
  \inheritanchorborder[from=rectangle]

  \anchor{center}{\pgfpointorigin}
  \anchor{north}{\northeast \pgf@x=0pt}
  \anchor{east}{\northeast \pgf@y=0pt}
  \anchor{south}{\southwest \pgf@x=0pt}
  \anchor{west}{\southwest \pgf@y=0pt}
  \anchor{north east}{\northeast}
  \anchor{north west}{\northeast \pgf@x=-\pgf@x}
  \anchor{south west}{\southwest}
  \anchor{south east}{\southwest \pgf@x=-\pgf@x}
  \anchor{text}{
    \pgfpointorigin
    \advance\pgf@x by -.5\wd\pgfnodeparttextbox%
    \advance\pgf@y by -.5\ht\pgfnodeparttextbox%
    \advance\pgf@y by +.5\dp\pgfnodeparttextbox%
  }

  \anchor{inputa}{
    \pgf@process{\northeast}%
    \pgf@x=-1\pgf@x%
    \pgf@y=.5\pgf@y%
  }
  \anchor{inputb}{
    \pgf@process{\northeast}%
    \pgf@x=-1\pgf@x%
    \pgf@y=.0\pgf@y%
  }
  \anchor{inputc}{
    \pgf@process{\northeast}%
    \pgf@x=-1\pgf@x%
    \pgf@y=-.5\pgf@y%
  }
  \backgroundpath{
    \pgfpathrectanglecorners{\southwest}{\northeast}

    \begingroup
    \tikzset{table node/port labels} % Use font from this style
    \tikz@textfont

    \pgf@anchor@table@inputa
    \pgftext[left,base,at={\pgfpoint{\pgf@x}{\pgf@y}},x=\pgfshapeinnerxsep]{\raisebox{-0.75ex}{I0}}
    
    \pgf@anchor@table@inputb
    \pgftext[left,base,at={\pgfpoint{\pgf@x}{\pgf@y}},x=\pgfshapeinnerxsep]{\raisebox{-0.75ex}{I1}}

    \pgf@anchor@table@inputc
    \pgftext[left,base,at={\pgfpoint{\pgf@x}{\pgf@y}},x=\pgfshapeinnerxsep]{\raisebox{-0.75ex}{\vdots{}}}

    \endgroup
  }
}

\pgfdeclareshape{view}{
  \savedanchor\northeast{%
    \pgfmathsetlength\pgf@x{\pgfshapeminwidth}%
    \pgfmathsetlength\pgf@y{\pgfshapeminheight}%
    \pgf@x=0.5\pgf@x
    \pgf@y=0.5\pgf@y
  }
  \savedanchor\southwest{%
    \pgfmathsetlength\pgf@x{\pgfshapeminwidth}%
    \pgfmathsetlength\pgf@y{\pgfshapeminheight}%
    \pgf@x=-0.5\pgf@x
    \pgf@y=-0.5\pgf@y
  }
  \inheritanchorborder[from=trapezium]

  \anchor{center}{\pgfpointorigin}
  \anchor{north}{\northeast \pgf@x=0pt}
  \anchor{east}{\northeast \pgf@y=0pt}
  \anchor{south}{\southwest \pgf@x=0pt}
  \anchor{west}{\southwest \pgf@y=0pt}
  \anchor{north east}{\northeast}
  \anchor{north west}{\northeast \pgf@x=-\pgf@x}
  \anchor{south west}{\southwest}
  \anchor{south east}{\southwest \pgf@x=-\pgf@x}
  \anchor{text}{
    \pgfpointorigin
    \advance\pgf@x by -.5\wd\pgfnodeparttextbox%
    \advance\pgf@y by -.5\ht\pgfnodeparttextbox%
    \advance\pgf@y by +.5\dp\pgfnodeparttextbox%
  }

  \anchor{inputa}{
    \pgf@process{\northeast}%
    \pgf@x=-1\pgf@x%
    \pgf@y=.5\pgf@y%
  }
  \anchor{inputb}{
    \pgf@process{\northeast}%
    \pgf@x=-1\pgf@x%
    \pgf@y=-.5\pgf@y%
  }
  \backgroundpath{
    \pgfpathrectanglecorners{\southwest}{\northeast}

    \begingroup
    \tikzset{view node/port labels} % Use font from this style
    \tikz@textfont

    \pgf@anchor@view@inputa
    \pgftext[left,base,at={\pgfpoint{\pgf@x}{\pgf@y}},x=\pgfshapeinnerxsep]{\raisebox{-0.75ex}{I0}}
    
    \pgf@anchor@view@inputb
    \pgftext[left,base,at={\pgfpoint{\pgf@x}{\pgf@y}},x=\pgfshapeinnerxsep]{\raisebox{-0.75ex}{I1}}

    \endgroup
  }
}

\pgfdeclareshape{bignode}{
  \savedanchor\northeast{%
    \pgfmathsetlength\pgf@x{\pgfshapeminwidth}%
    \pgfmathsetlength\pgf@y{\pgfshapeminheight}%
    \pgf@x=0.5\pgf@x
    \pgf@y=0.5\pgf@y
  }
  \savedanchor\southwest{%
    \pgfmathsetlength\pgf@x{\pgfshapeminwidth}%
    \pgfmathsetlength\pgf@y{\pgfshapeminheight}%
    \pgf@x=-0.5\pgf@x
    \pgf@y=-0.5\pgf@y
  }
  \inheritanchorborder[from=rectangle]

  \anchor{center}{\pgfpointorigin}
  \anchor{north}{\northeast \pgf@x=0pt}
  \anchor{east}{\northeast \pgf@y=0pt}
  \anchor{south}{\southwest \pgf@x=0pt}
  \anchor{west}{\southwest \pgf@y=0pt}
  \anchor{north east}{\northeast}
  \anchor{north west}{\northeast \pgf@x=-\pgf@x}
  \anchor{south west}{\southwest}
  \anchor{south east}{\southwest \pgf@x=-\pgf@x}
  \anchor{text}{
    \pgfpointorigin
    \advance\pgf@x by -.5\wd\pgfnodeparttextbox%
    \advance\pgf@y by -.5\ht\pgfnodeparttextbox%
    \advance\pgf@y by +.5\dp\pgfnodeparttextbox%
  }

  \anchor{inputa}{
    \pgf@process{\northeast}%
    \pgf@x=-1\pgf@x%
    \pgf@y=.5\pgf@y%
  }
  \anchor{inputb}{
    \pgf@process{\northeast}%
    \pgf@x=-1\pgf@x%
    \pgf@y=.1\pgf@y%
  }
  \anchor{inputi}{
    \pgf@process{\northeast}%
    \pgf@x=-1\pgf@x%
    \pgf@y=-.3\pgf@y%
  }
  \anchor{inputn}{
    \pgf@process{\northeast}%
    \pgf@x=-1\pgf@x%
    \pgf@y=-.6\pgf@y%
  }
  \anchor{outputa}{
    \pgf@process{\northeast}%
    \pgf@y=.5\pgf@y%
  }
  \anchor{outputb}{
    \pgf@process{\northeast}%
    \pgf@y=.1\pgf@y%
  }
  \anchor{outputi}{
    \pgf@process{\northeast}%
    \pgf@y=-.3\pgf@y%
  }
  \anchor{outputn}{
    \pgf@process{\northeast}%
    \pgf@y=-.6\pgf@y%
  }
  \anchor{confa}{
    \pgf@process{\northeast}%
    \pgf@x=-.5\pgf@x%
  }
  \anchor{confb}{
    \pgf@process{\northeast}%
    \pgf@x=-.1\pgf@x%
  }
  \anchor{confi}{
    \pgf@process{\northeast}%
    \pgf@x=.2\pgf@x%
  }
  \anchor{confn}{
    \pgf@process{\northeast}%
    \pgf@x=.5\pgf@x%
  }
  \backgroundpath{
    \pgfpathrectanglecorners{\southwest}{\northeast}

    \begingroup
    \tikzset{bignode node/port labels} % Use font from this style
    \tikz@textfont

    \pgf@anchor@bignode@inputa
    \pgftext[left,base,at={\pgfpoint{\pgf@x}{\pgf@y}},x=\pgfshapeinnerxsep]{\raisebox{-0.75ex}{Input \#0}}

    \pgf@anchor@bignode@inputb
    \pgftext[left,base,at={\pgfpoint{\pgf@x}{\pgf@y}},x=\pgfshapeinnerxsep]{\raisebox{-0.75ex}{Input \#1}}
    
    \pgf@anchor@bignode@inputi
    \pgftext[left,base,at={\pgfpoint{\pgf@x}{\pgf@y}},x=\pgfshapeinnerxsep]{\raisebox{-0.75ex}{\vdots}}

    \pgf@anchor@bignode@inputn
    \pgftext[left,base,at={\pgfpoint{\pgf@x}{\pgf@y}},x=\pgfshapeinnerxsep]{\raisebox{-0.75ex}{Input \#n}}

    \pgf@anchor@bignode@outputa
    \pgftext[right,base,at={\pgfpoint{\pgf@x}{\pgf@y}},x=-\pgfshapeinnerxsep]{\raisebox{-.75ex}{Output \#0}}

    \pgf@anchor@bignode@outputb
    \pgftext[right,base,at={\pgfpoint{\pgf@x}{\pgf@y}},x=-\pgfshapeinnerxsep]{\raisebox{-.75ex}{Output \#1}}

    \pgf@anchor@bignode@outputi
    \pgftext[right,base,at={\pgfpoint{\pgf@x}{\pgf@y}},x=-\pgfshapeinnerxsep]{\raisebox{-.75ex}{\vdots}}

    \pgf@anchor@bignode@outputn
    \pgftext[right,base,at={\pgfpoint{\pgf@x}{\pgf@y}},x=-\pgfshapeinnerxsep]{\raisebox{-.75ex}{Output \#o}}

    \pgf@anchor@bignode@confa
    \pgftext[top,at={\pgfpoint{\pgf@x}{\pgf@y}},y=-\pgfshapeinnerysep]{Config \#1}

    \pgf@anchor@bignode@confb
    \pgftext[top,at={\pgfpoint{\pgf@x}{\pgf@y}},y=-\pgfshapeinnerysep]{Config \#2}

    \pgf@anchor@bignode@confi
    \pgftext[top,at={\pgfpoint{\pgf@x}{\pgf@y}},y=-\pgfshapeinnerysep]{\dots}

    \pgf@anchor@bignode@confn
    \pgftext[top,at={\pgfpoint{\pgf@x}{\pgf@y}},y=-\pgfshapeinnerysep]{Config \#m}

    \endgroup
  }
}

\pgfdeclareshape{smallnode}{
  \savedanchor\northeast{%
    \pgfmathsetlength\pgf@x{\pgfshapeminwidth}%
    \pgfmathsetlength\pgf@y{\pgfshapeminheight}%
    \pgf@x=0.5\pgf@x
    \pgf@y=0.5\pgf@y
  }
  \savedanchor\southwest{%
    \pgfmathsetlength\pgf@x{\pgfshapeminwidth}%
    \pgfmathsetlength\pgf@y{\pgfshapeminheight}%
    \pgf@x=-0.5\pgf@x
    \pgf@y=-0.5\pgf@y
  }
  \inheritanchorborder[from=rectangle]

  \anchor{center}{\pgfpointorigin}
  \anchor{north}{\northeast \pgf@x=0pt}
  \anchor{east}{\northeast \pgf@y=0pt}
  \anchor{south}{\southwest \pgf@x=0pt}
  \anchor{west}{\southwest \pgf@y=0pt}
  \anchor{north east}{\northeast}
  \anchor{north west}{\northeast \pgf@x=-\pgf@x}
  \anchor{south west}{\southwest}
  \anchor{south east}{\southwest \pgf@x=-\pgf@x}
  \anchor{text}{
    \pgfpointorigin
    \advance\pgf@x by -.5\wd\pgfnodeparttextbox%
    \advance\pgf@y by -.5\ht\pgfnodeparttextbox%
    \advance\pgf@y by +.5\dp\pgfnodeparttextbox%
  }

    \anchor{inputa}{
    \pgf@process{\northeast}%
    \pgf@x=-1\pgf@x%
    \pgf@y=.4\pgf@y%
  }
  \anchor{inputb}{
    \pgf@process{\northeast}%
    \pgf@x=-1\pgf@x%
    \pgf@y=-.4\pgf@y%
  }
  \anchor{outputa}{
    \pgf@process{\northeast}%
    \pgf@y=.4\pgf@y%
  }
  \anchor{outputb}{
    \pgf@process{\northeast}%
    \pgf@y=-.4\pgf@y%
  }
  \anchor{confa}{
    \pgf@process{\northeast}%
    \pgf@x=-.5\pgf@x%
  }
  \anchor{confb}{
    \pgf@process{\northeast}%
    \pgf@x=.5\pgf@x%
  }
  \backgroundpath{
    \pgfpathrectanglecorners{\southwest}{\northeast}

    \begingroup
    \tikzset{smallnode node/port labels} % Use font from this style
    \tikz@textfont

    \pgf@anchor@smallnode@inputa
    \pgftext[left,base,at={\pgfpoint{\pgf@x}{\pgf@y}},x=\pgfshapeinnerxsep]{\raisebox{-0.75ex}{}}
    
    \pgf@anchor@smallnode@inputb
    \pgftext[left,base,at={\pgfpoint{\pgf@x}{\pgf@y}},x=\pgfshapeinnerxsep]{\raisebox{-0.75ex}{}}

    \pgf@anchor@smallnode@outputa
    \pgftext[right,base,at={\pgfpoint{\pgf@x}{\pgf@y}},x=-\pgfshapeinnerxsep]{\raisebox{-.75ex}{}}

    \pgf@anchor@smallnode@outputb
    \pgftext[right,base,at={\pgfpoint{\pgf@x}{\pgf@y}},x=-\pgfshapeinnerxsep]{\raisebox{-.75ex}{}}

    \pgf@anchor@smallnode@confa
    \pgftext[top,at={\pgfpoint{\pgf@x}{\pgf@y}},y=-\pgfshapeinnerysep]{}

    \pgf@anchor@smallnode@confb
    \pgftext[top,at={\pgfpoint{\pgf@x}{\pgf@y}},y=-\pgfshapeinnerysep]{}

    \endgroup
  }
}

\pgfdeclareshape{function}{
  \savedanchor\northeast{%
    \pgfmathsetlength\pgf@x{\pgfshapeminwidth}%
    \pgfmathsetlength\pgf@y{\pgfshapeminheight}%
    \pgf@x=0.5\pgf@x
    \pgf@y=0.5\pgf@y
  }
  \savedanchor\southwest{%
    \pgfmathsetlength\pgf@x{\pgfshapeminwidth}%
    \pgfmathsetlength\pgf@y{\pgfshapeminheight}%
    \pgf@x=-0.5\pgf@x
    \pgf@y=-0.5\pgf@y
  }
  \inheritanchorborder[from=rectangle]

  \anchor{center}{\pgfpointorigin}
  \anchor{north}{\northeast \pgf@x=0pt}
  \anchor{east}{\northeast \pgf@y=0pt}
  \anchor{south}{\southwest \pgf@x=0pt}
  \anchor{west}{\southwest \pgf@y=0pt}
  \anchor{north east}{\northeast}
  \anchor{north west}{\northeast \pgf@x=-\pgf@x}
  \anchor{south west}{\southwest}
  \anchor{south east}{\southwest \pgf@x=-\pgf@x}
  \anchor{text}{
    \pgfpointorigin
    \advance\pgf@x by -.5\wd\pgfnodeparttextbox%
    \advance\pgf@y by -.5\ht\pgfnodeparttextbox%
    \advance\pgf@y by +.5\dp\pgfnodeparttextbox%
  }

  \anchor{inputa}{
    \pgf@process{\northeast}%
    \pgf@x=-1\pgf@x%
    \pgf@y=.5\pgf@y%
  }
  \anchor{outputa}{
    \pgf@process{\northeast}%
    \pgf@y=.5\pgf@y%
  }
  \anchor{outputb}{
    \pgf@process{\northeast}%
    \pgf@y=0\pgf@y%
  }
  \anchor{outputc}{
    \pgf@process{\northeast}%
    \pgf@y=-.5\pgf@y%
  }
  \backgroundpath{
    \pgfpathrectanglecorners{\southwest}{\northeast}

    \begingroup
    \tikzset{function node/port labels} % Use font from this style
    \tikz@textfont

    \pgf@anchor@function@inputa
    \pgftext[left,base,at={\pgfpoint{\pgf@x}{\pgf@y}},x=\pgfshapeinnerxsep]{\raisebox{-0.75ex}{I0}}

    \pgf@anchor@function@outputa
    \pgftext[right,base,at={\pgfpoint{\pgf@x}{\pgf@y}},x=-\pgfshapeinnerxsep]{\raisebox{-.75ex}{O0}}
    
    \pgf@anchor@function@outputb
    \pgftext[right,base,at={\pgfpoint{\pgf@x}{\pgf@y}},x=-\pgfshapeinnerxsep]{\raisebox{-.75ex}{O1}}

    \pgf@anchor@function@outputc
    \pgftext[right,base,at={\pgfpoint{\pgf@x}{\pgf@y}},x=-\pgfshapeinnerxsep]{\raisebox{-.75ex}{\vdots{}}}

    \endgroup
  }
}

\tikzset{add font/.code={\expandafter\def\expandafter\tikz@textfont\expandafter{\tikz@textfont#1}}} 

\tikzset{decision node/port labels/.style={font=\sffamily\scriptsize}}
\tikzset{every decision node/.style={draw,minimum width=3cm,minimum 
height=2cm,very thick,inner sep=1mm,outer sep=0pt,cap=round,add 
font=\sffamily}}

\tikzset{nodea node/port labels/.style={font=\sffamily\scriptsize}}
\tikzset{every nodea node/.style={draw,minimum width=3cm,minimum 
height=2cm,very thick,inner sep=1mm,outer sep=0pt,cap=round,add 
font=\sffamily}}

\tikzset{nodeb node/port labels/.style={font=\sffamily\scriptsize}}
\tikzset{every nodeb node/.style={draw,minimum width=3cm,minimum 
height=2cm,very thick,inner sep=1mm,outer sep=0pt,cap=round,add 
font=\sffamily}}

\tikzset{nodec node/port labels/.style={font=\sffamily\scriptsize}}
\tikzset{every nodec node/.style={draw,minimum width=3cm,minimum 
height=2cm,very thick,inner sep=1mm,outer sep=0pt,cap=round,add 
font=\sffamily}}

\tikzset{noded node/port labels/.style={font=\sffamily\scriptsize}}
\tikzset{every noded node/.style={draw,minimum width=3cm,minimum 
height=2cm,very thick,inner sep=1mm,outer sep=0pt,cap=round,add 
font=\sffamily}}

\tikzset{nodee node/port labels/.style={font=\sffamily\scriptsize}}
\tikzset{every nodee node/.style={draw,minimum width=3cm,minimum 
height=2cm,very thick,inner sep=1mm,outer sep=0pt,cap=round,add 
font=\sffamily}}

\tikzset{nodef node/port labels/.style={font=\sffamily\scriptsize}}
\tikzset{every nodef node/.style={draw,minimum width=3cm,minimum 
height=2cm,very thick,inner sep=1mm,outer sep=0pt,cap=round,add 
font=\sffamily}}

\tikzset{nodeg node/port labels/.style={font=\sffamily\scriptsize}}
\tikzset{every nodeg node/.style={draw,minimum width=3cm,minimum 
height=2cm,very thick,inner sep=1mm,outer sep=0pt,cap=round,add 
font=\sffamily}}

\tikzset{summary node/port labels/.style={font=\sffamily\scriptsize}}
\tikzset{every summary node/.style={draw,minimum width=4cm,minimum 
height=2cm,very thick,inner sep=1mm,outer sep=0pt,cap=round,add 
font=\sffamily}}

\tikzset{table node/port labels/.style={font=\sffamily\scriptsize}}
\tikzset{every table node/.style={draw,minimum width=3cm,minimum 
height=2cm,very thick,inner sep=1mm,outer sep=0pt,cap=round,add 
font=\sffamily}}

\tikzset{view node/port labels/.style={font=\sffamily\scriptsize}}
\tikzset{every view node/.style={draw,minimum width=3cm,minimum 
height=2cm,very thick,inner sep=1mm,outer sep=0pt,cap=round,add 
font=\sffamily}}

\tikzset{function node/port labels/.style={font=\sffamily\scriptsize}}
\tikzset{every function node/.style={draw,minimum width=3cm,minimum 
height=2cm,very thick,inner sep=1mm,outer sep=0pt,cap=round,add 
font=\sffamily}}

\tikzset{bignode node/port labels/.style={font=\sffamily\scriptsize}}
\tikzset{every bignode node/.style={draw,minimum width=7cm,minimum 
height=3cm,very thick,inner sep=1mm,outer sep=0pt,cap=round,add 
font=\sffamily}}

\tikzset{smallnode node/port labels/.style={font=\sffamily\footnotesize}}
\tikzset{every smallnode node/.style={draw,minimum width=3cm,minimum 
height=1cm,very thick,inner sep=1mm,outer sep=0pt,cap=round,add 
font=\sffamily}}

\makeatother

\tikzstyle{startstop} = [rectangle, rounded corners, minimum width=0.2cm, minimum height=0.2cm, text centered, text width=1cm, font=\scriptsize, draw=black]
\tikzstyle{io_small} = [trapezium, trapezium left angle=70, trapezium right angle=110, minimum width=0.2cm, minimum height=0.2cm, text centered, text width=1cm, font=\scriptsize, draw=black]
\tikzstyle{io} = [trapezium, trapezium left angle=70, trapezium right angle=110, minimum width=0.5cm, minimum height=0.7cm, text centered, text width=1.5cm, font=\scriptsize, draw=black]
\tikzstyle{process} = [rectangle, minimum width=0.2cm, minimum height=0.2cm, text centered, text width=1.5cm, font=\scriptsize, draw=black]
\tikzstyle{decision} = [diamond, minimum width=0.2cm, minimum height=0.2cm, text centered, text width=1.2cm, font=\scriptsize, draw=black]
\tikzstyle{arrow} = [thick,->,>=stealth]

\tikzset{three sided/.style={
        draw=none,
        append after command={
            [shorten <= -0.5\pgflinewidth]
            ([shift={(-1.5\pgflinewidth,-0.5\pgflinewidth)}]\tikzlastnode.north east)
        edge([shift={( 0.5\pgflinewidth,-0.5\pgflinewidth)}]\tikzlastnode.north west) 
            ([shift={( 0.5\pgflinewidth,-0.5\pgflinewidth)}]\tikzlastnode.north west)
        edge([shift={( 0.5\pgflinewidth,+0.5\pgflinewidth)}]\tikzlastnode.south west)            
            ([shift={( 0.5\pgflinewidth,+0.5\pgflinewidth)}]\tikzlastnode.south west)
        edge([shift={(-1.0\pgflinewidth,+0.5\pgflinewidth)}]\tikzlastnode.south east)
        }
    }
}

\makeatletter
\let\saved@longtable\longtable
\long\def\foo#1\LT@err#2#3#4!!{\def\longtable{#1#4}}
\expandafter\foo\longtable!!

\long\def\foo#1\@outputpage#2\@outputpage#3!!{%
\def\LT@output{#1\@opcol#2\@opcol#3}}
\expandafter\foo\LT@output!!
\makeatother

\begin{document}

% Acronyms.
\newacronym{tsg}{TSG}{troubleshooting graph}
\newacronym{tramp}{TRAMP}{Transparent Remote Access, Multiple Protocols}
\newacronym{sdn}{SDN}{software defined network}
\newacronym{dpid}{DPID}{dapatapath identifier}
\newacronym{gdb}{GDB}{GNU Debugger}
\newacronym{ovs}{OVS}{Open vSwitch}
\newacronym{arp}{ARP}{Address Resolution Protocol}

\title{A Little Less Interaction, A Little More Action: \\ 
       A Modular Framework for Network Troubleshooting}

\author[1, *]{\pelle{}}
\author[1]{\nemeth{}}
\author[2]{\gulyas{}}
\affil[1]{Budapest University of Technology and Economics, Hungary, HSNLab, Dept. of Telecommunications and Media Informatics}
\affil[2]{Budapest University of Technology and Economics, Hungary, HSNLab, Dept. of Telecommunications and Media Informatics and MTA-BME Information Systems Research Group and was supported by the J\'anos Bolyai Fellowship of the Hungarian Academy of Sciences}
\affil[*]{Corresponding author: \pelle{} (pelle@tmit.bme.hu)}

\renewcommand\Authands{ and }

% make the title area
\maketitle

\begin{abstract}
  An ideal network troubleshooting system would be an almost fully
  automated system, monitoring the whole network at once, feeding the
  results to a knowledge-based decision making system that suggests
  actions to the operator or corrects the failure
  automatically. Reality is quite the contrary: operators separated in
  their offices try to track down complex networking failures in their
  own way, which is generally a long sequence of manually edited
  parallel shell commands (mostly ping, traceroute, route, iperf,
  ofctl etc.). This process requires operators to be ``masters of
  complexity'' (which they often are) and continuous interaction.  In
  this paper we aim at narrowing this huge gap between vision and
  reality by introducing a modular framework capable of 
  \begin{enumerate*}[label=\emph{(\roman*)}]
  \item formalizing troubleshooting processes as the concatenation of
    executable functions [called \glspl{tsg}],
  \item executing these graphs via an interpreter,
  \item evaluating and navigating between the outputs of the functions and
  \item sharing troubleshooting know-hows in a formalized manner.
  \end{enumerate*}  
\end{abstract}

\section{Introduction}

Troubleshooting a communication network was never an easy
problem. Finding causes of errors and failures, tracking down
misconfigurations in the increasingly complex interconnection networks
of heterogeneous networking devices is quite a challenge. What is
more, the prevalence of increasingly complex software components, due
to the upcoming \glspl{sdn}, adds distributed software debugging as an
additional issue to deal with. To cope with this increasing
complexity, the networking research community suggests the use of
knowledge-based decision support together with the standard network
monitoring and diagnostic tools, and the conversion of troubleshooting
into a highly automated process.  Reality seems to reside very far
away from this vision. Real operators tend to use the most basic
diagnostic tools for monitoring the network, and rely on their own
brilliance and programming skills when digging out the root causes of
errors in an ad-hoc manner from the reports of these tools. Even if
this approach works well in practice, it requires extremely skilled
operators who can keep in mind all the details of the network under
scrutiny and their continuous interaction usually is wasted on
rummaging in the logs of the tools used by them.

As we see, the reason for this huge gap between the ideas and reality
is threefold. First, there is no usable, implementation oriented
formal description of the troubleshooting processes. Secondly, there
is no platform capable of executing formally defined troubleshooting
processes while giving prompt and systematic access to the outputs of
the used tools. Finally, there is no existing platform that could
integrate existing troubleshooting tools and decision support
methodologies in a flexible manner. In lack of formalism and
integrated execution platform, operators cannot share and re-use each
others troubleshooting know-hows in a structured way, thus knowledge
is not accumulated but remains sporadic as operators treat every
specific failure in their own ad-hoc way.

The purpose of this paper is narrowing the gap between troubleshooting
visions and real life solutions. In this respect, our contribution
will be threefold. First, we propose a formalization of
troubleshooting processes in the form of \glsentryfullpl{tsg},
which let operators describe the steps of tracking down a specific
network failure in a structural manner with a very small effort
compared to the implementation of it. Once created, \glspl{tsg} can make
their solutions ready-to-share and re-usable. We also define a
language for a text-based representation of \glspl{tsg}.  Secondly, we
propose a modular execution framework capable of running \glspl{tsg} and
giving on demand fast semantic navigation among the outputs of the
tools used in the troubleshooting process. Finally, we present a
complete prototype system (called \emph{\epoxide{}}) capable of defining,
executing and analyzing \glspl{tsg}. Case studies using \epoxide{} are also
given.

The rest of our paper is structured as follows: in
Section~\ref{sec:sota} we give a brief overview on the related work in
both literature and practice. Section~\ref{sec:principles} lists the
principles of our proposed modular troubleshooting framework, followed
by the illustration of its operation over an everyday example in
Section~\ref{sec:example}. Section~\ref{sec:implementation} presents
the fundamentals of our prototype, \epoxide{}, which is complemented
with some illustrative case studies in
Section~\ref{sec:case}. Finally, we conclude the paper and give
directions for future works in Section~\ref{sec:conclusion}.

\section{State of the Art in Network Troubleshooting}
\label{sec:sota}
From the great volume of related literature we highlight here the two
main constituents of troubleshooting systems. The first is clearly the
area of \emph{network monitoring and diagnostic tools}, of which main
purpose is to seek for symptoms of specific failures.  The palette is
very broad here, ranging from the most basic tools---like ping,
traceroute, tcpdump, netstat, nmap~\cite{sloan2001network} or
\gls{gdb}---, through monitoring protocols---such as SNMP and
RMON~\cite{sloan2001network}---, configuration files and
analyzers---such as Splat~\cite{abrahamson2003splat}---, performance
measurement tools---such as iperf~\cite{sloan2001network}---and packet
analyzers---like Wireshark---, to the more complex ones---such as
NetFlow, HSA~\cite{kazemian2013real,kazemian2012header} and
ATPG~\cite{zeng2012automatic}.  On top of these, \gls{sdn} specific
tools have added a whole new segment targeted to investigate specific
parts of the architecture.  Tools such as
Anteater~\cite{Mai:2011:DDP:2018436.2018470},
OFRewind~\cite{wundsam2011ofrewind}, NetSight~\cite{handigol2014know},
VeriFlow~\cite{khurshid2012veriflow}, NICE~\cite{canini2012nice},
SOFT~\cite{kuzniar2012soft}, FORTNOX~\cite{porras2012security} and
OFTEN~\cite{kuzniar2012often} all fill a niche in \gls{sdn}
troubleshooting.

One level up, the output symptoms of these tools can be aggregated and
fed into different \emph{automatic reasoning solutions}. The first
representatives of these were created as early as the second half of
the 1980s~\cite{mathonet1987dantes} targeting the discovery of
failures in telecommunication networks.  Early on, rule-based methods
were used to resolve issues by using \emph{if--then}
statements~\cite{hitson1988knowledge-based}.  Later case-based
reasoning~\cite{lewis1993case-based} and
model-based~\cite{jakobson1995real-time} methods were developed.  The
former utilized a collection of previous cases as a basis for failure
analysis, while the latter used models of structural and functional
behavior to reason about network issues.  Fault-symptom
graphs~\cite{reali2008ngn_ims} and dependency or causality
graphs~\cite{lu2010self-healing, lu2011self-diagnosis} introduced the
concept of tracking failures using graphs that created connections
between symptoms, detection and root causes.  This concept led to the
application of Bayesian networks~\cite{charniak1991bayesian,
  khanafer2008automated} where belief---in the most probable failure
root cause---propagation is based on a probabilistic model.

\subsection{What We See in Current Practice}

Despite the readily available set of advanced troubleshooting tools
and decision support mechanisms, operators seem to use the most
rudimentary tools (like ping, traceroute, tcpdump etc.) while they
completely rely on their minds as a knowledge-base. For testing this,
we conducted an in-house survey querying which type of problems local
administrators run into most frequently and what network
troubleshooting tools they use most commonly.  The results that we
found were completely in accordance with those outlined
in~\cite{zeng2012survey}.  Most problems were caused by connectivity
issues that arose from a variety of reasons ranging from hardware
failures to configuration changes that became necessary due to
security issues.  Used troubleshooting tools show similarities also:
mostly simple task specific tools are utilized, in certain cases
combining them in a script to explore typical failures.  Network
information is usually stored in simple spreadsheets and proprietary
monitoring or troubleshooting tools are used only when they have a low
cost---or are preferably free.  We found that automatic tools are less
frequently used and manual troubleshooting dominates problem solving.

To get a deeper sense of the process, think of an operator logging
into different devices and running heterogeneous software tools
(e.g. the tools listed above) to analyze the problem.  Network
conditions are monitored simultaneously in multiple shells and after a
painful rummaging in the tools' outputs, new commands are evoked on
new devices. We argue that this approach has its own merits and
drawbacks. It is extremely flexible as the operator uses the tool of
her choosing in the way and logic she sees fit. On the negative side,
it is quite ineffective, unorganized and anything but re-usable.  The
operator's time is spent mostly on filtering and finding the
correlation between the different outputs and keeping in mind the
mapping between different shells and devices.

\section{Design Principles of a Modular Framework for Network Troubleshooting}
\label{sec:principles}

Instead of proposing a new troubleshooting tool or another decision
support mechanism, we suggest here a framework\footnote{Our initial
  research and implementation of a subset of current framework
  functionality is discussed in~\cite{pelle2015onetool}.} capable of
\emph{combining} existing (and future) special-purpose tools and
reasoning methodologies in a modular fashion.  Our concept builds on
the observation that operators combine different troubleshooting tools
to find out the root cause of a network issue.  In the following
sections, we go through the main notions that we use to describe such
troubleshooting processes and the fundamentals of our framework
capable of executing \glsentryfullpl{tsg}.

\subsection{Nodes: Wrappers Around Troubleshooting Tools}

\begin{figure}[!t]
  \centering
  \begin{tikzpicture}[font=\sffamily]

    \node (i0) [coordinate, xshift=-4.5cm, yshift=0.75cm] {};
    \node (i1) [coordinate, below of=i0, yshift=0.4cm] {};
    \node (in) [coordinate, below of=i1, yshift=-0.05cm] {};

    \node (c0) [coordinate, xshift=-1.75cm, yshift=2.5cm] {};
    \node (c1) [coordinate, right of=c0, xshift=0.4cm] {};
    \node (cn) [coordinate, right of=c1, xshift=1.1cm] {};

    \node (o0) [coordinate, xshift=4.5cm, yshift=0.75cm] {};
    \node (o1) [coordinate, below of=o0, yshift=0.4cm] {};
    \node (on) [coordinate, below of=o1, yshift=-0.05cm] {};

    \node [bignode] (wrapper) {};
    \node [smallnode, yshift=-0.5cm] (tool) {Wrapped tool};
    
    \draw [->] (.6, 1.3) -- (.6, .5) -- (-.75, .5) -- (tool.confa);
    \draw [->] (.75, 1.3) -- (tool.confb);
    \draw [->] (-3.3, -.3) -- (tool.inputa);
    \draw [->] (-3.3, -.5) -- (-2.5, -.5) -- (-2.5, -.7) -- (tool.inputb);
    \draw [->] (tool.outputa)  -- (3.3, -.3);
    \draw [->] (tool.outputb) -- (2.5, -.7) -- (2.5, -.5) -- (3.3, -.5);

    \draw [->] (i0) -- (wrapper.inputa);
    \draw [->] (i1) -- (wrapper.inputb);
    \draw [->] (in) -- (wrapper.inputn);
    
    \draw [<-] (o0) -- (wrapper.outputa);
    \draw [<-] (o1) -- (wrapper.outputb);
    \draw [<-] (on) -- (wrapper.outputn);

    \draw [->] (c0) -- (wrapper.confa);
    \draw [->] (c1) -- (wrapper.confb);
    \draw [->] (cn) -- (wrapper.confn);

  \end{tikzpicture}
  \caption{A conceptual node.}
  \label{fig:node}
\end{figure}
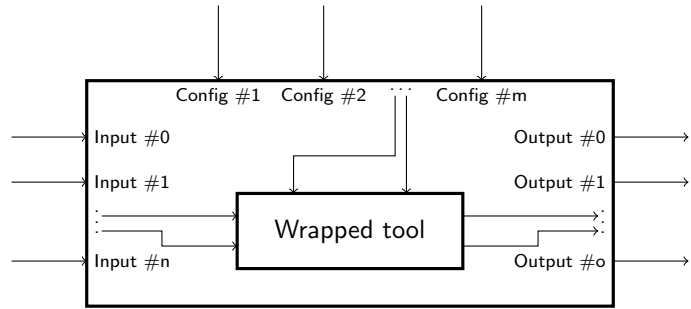

The first thing we need is an abstraction that incorporates the basic
elements of a troubleshooting process.  Thus we define our
\emph{nodes} as wrappers around troubleshooting tools or smaller,
processing functions.  Nodes are considered as black boxes that hide
their internal operation from the operator (see
Fig.~\ref{fig:node}). Operators have three types of interfaces for
communicating with nodes: inputs, on which the nodes execute
operations (e.g.~a text stream to process or clock ticks),
configuration arguments and outputs for relaying information.

The outputs can relay the exact output of the wrapped tool or provide
extra processing before generating their results.  Nodes are also
responsible for providing documentation for these interfaces.

Nodes have three stages of their life cycle: initialization, execution
and termination.  The initialization stage is where the environment
setup of the node is done, including resource allocation and initial
configuration.  Nodes enter this stage of their life cycle only once.
At the execution stage, nodes read the data arriving on their inputs and
start the wrapped process or function. Analysis or modification on the
wrapped tool's output is also performed here. The node is constantly in
this stage when it has been initialized but not yet been terminated.
Finally, the node reaches the termination stage when it is being
stopped.  This stage is responsible for clearing up allocated
resources and terminating wrapped processes.

We identified two basic node types: processing nodes and display
nodes.  The first interprets its inputs and---after processing---it
relays the results on its outputs.  The second type only interprets
input data and is responsible for creating a graphical representation
for it, e.g.~in the form of a chart or a graph.

For an example, consider a basic node wrapping the \texttsg{traceroute}
tool. The node outputs the trace (optionally from a remote machine) to
a specific host every time we call it.  Let's define first the node's
interfaces:

\begin{itemize}
\item Inputs: only a single input is needed for receiving an enabling
  signal.  Each time this input changes, the node should call the
  \texttsg{traceroute} command.
\item Configuration arguments: we can simply choose to make these the
  same as the command line arguments of the \texttsg{traceroute}
  command. If we want to run traceroute on a remote machine, we can
  add another argument to specify the remote host.
\item Outputs: the node should output the results returned by the
  \texttsg{traceroute} call on its first output.  On the second
  output, it shows the last hop on the way to the specified target.
\end{itemize}

In the initialization stage, when requested, the node sets up
connection to the remote host where \texttsg{traceroute} should be
run. At execution time, it calls the \texttsg{traceroute} command and
relays its unmodified result to the first output while analyzing it.
If the analysis found the last hop, the node sends this information to
its second output. In the termination stage, the ongoing processes are
stopped and the connection to the remote host is closed, if it had
been set up.

\subsection{Edges: Accessible Data Transfer}

We use \emph{edges} to describe the connections between nodes.
Besides specifying the nodes to connect, the main feature required
from edges is to provide accessibility for node outputs: information
over the edges is observable and modifiable on demand by the operator.
When access is provided to edges, the operator can see what is
happening in the troubleshooting process on the lowest levels.  When
historical data is available on the edges, the operator can backtrack
how network conditions changed during runtime.  Modifiable edges
provide the additional benefit of direct operator interaction with
nodes, which is helpful for instant testing purposes.  Let's take the
example when our troubleshooting scenario is described as a path graph
and the first node infrequently outputs data.  In this case, the rest
of the nodes are also executed rarely and it is quite hard to test
whether our graph works as expected.  When having the option of
modifying edge content, we can easily inject test data on the first
edge instead of creating a test node that would supply the same data.
This method also helps channeling---otherwise unobtainable---data into
our system from the network environment.

\subsection{The Troubleshooting Graph}

Building on the concept of wrapper nodes and accessible edges, a
\gls{tsg} can be created that is able to describe troubleshooting
processes as series of tools and transformations.  The top part of
Fig.~\ref{fig:node-example} depicts a simple \gls{tsg} that
queries, in every five seconds, an SDN controller---accessible on the
local machine---for topology information and displays it using a graph
visualization node.  A special grouping node, the \texttsg{View}, creates a
grouping consisting of a node and an output, in order for these to be
displayed together.

Besides the simple concatenation of nodes, creating branches is
possible through special purpose \emph{decision} nodes.  These nodes
are processing nodes able to analyze incoming data and match against a
specific criteria set.  Such nodes can provide generalized decision
making apparatus that can combine results arriving from different
nodes and implement arbitrary decision functions to analyze and
evaluate them.  For example, such a function can validate the output
of the \texttsg{ifconfig} tool and decide whether or not each
interface is configured correctly.  Another example can be a function
that accepts numerical inputs and computes a weighted combination of
them, like nodes do in a Bayesian or neural network.  We do not
consider these decision functions as part of our tool, they can be
added to the framework by operators or third parties in a modular
fashion.  We note that a large collection of decision functions can
significantly shorten the creation time of effective branching
\glspl{tsg}.

\newsavebox{\shortexample}
\begin{lrbox}{\shortexample}
\begin{minipage}[t]{\linewidth}
\begin{lstlisting}[language=C++, showstringspaces=false,
    basicstyle=\ttfamily\footnotesize, columns=fullflexible, numbers=left,
    xleftmargin=2em, framexleftmargin=2em, escapeinside={(*@}{@*)}]
Clock(5) -> t :: Topology-SDN(localhost)(*@\label{lst:node-example-l1}@*)
-> Graph() --> view;(*@\label{lst:node-example-l2}@*)
t[0] -> [1]view;(*@\label{lst:node-example-l3}@*)
\end{lstlisting}
\end{minipage}
\end{lrbox}

\begin{figure}
  \centering
  \subfloat{
    \resizebox{9cm}{3.5cm}{
      \begin{tikzpicture}[node distance=5cm, font=\sffamily]
        
        \node [shape=nodea] (clock) {Clock(5)};
        
        \node [shape=nodeb, right of=clock] (topology) {\begin{tabular}{c}Topology-SDN\\(localhost)\end{tabular}};
        
        \draw [->, line width=.3mm] (clock.outputa) -- (topology.inputa);
        
        \node [shape=nodef, right of=topology, yshift=1.5cm] (graph) {Graph()};
        
        \draw [->, line width=.3mm] (topology.outputa) -- (7.5, 0.5) -- (7.5, 2) -- (graph.inputa);
        
        \node [shape=view, right of=topology, yshift=-1.5cm] (view) {View()};
        
        \draw [->, line width=.3mm] (graph) -- (10, 0) -- (8, 0) -- (8, -1) -- (view.inputa);
        
        \draw [->, line width=.3mm] (7.5, 0.5) -- (7.5, -2) -- (view.inputb);
        
      \end{tikzpicture}}
    \label{fig_first_case}}
  \hfill
  \subfloat{\usebox{\shortexample}}
  \caption{A simple \gls{tsg} example (top) and its description
    (bottom).}
  \label{fig:node-example}
\end{figure}

\subsubsection{A Language for Describing \glspl{tsg}}

For a text-based representation of TSGs, we define a simple
Click-inspired \cite{kohler2000click} description language.  Such
language fits perfectly to our concept as we look at nodes as black
boxes that have inputs, outputs and configuration arguments, which the
language supports by default.  Port-based explicit node linking is
also a feature that we make good use of.  Nodes can be defined by
assigning an instance name to a wrapper node using the \scope{}
operator (see line~\ref{lst:node-example-l1} of
Fig.~\ref{fig:node-example}).  For simplicity, the instance name can
be omitted if the node is not referenced in the code later on.
Configuration arguments follow the node instance assignment in
parentheses (see line~\ref{lst:node-example-l1}).  Nodes can be linked
with edges by linking expressions, using the \link{} linking operator.
Line~\ref{lst:node-example-l3} of the example shows a port-based
explicit linking of the topology and view nodes. A linking operator
always has the list of node outputs on its left side and the list of
inputs on its right side, thus multiple edges can be created between
two nodes in a single expression.  Outputs and inputs are always
referred to with their zero-based indices.  In case only the \nth{0}
input or output appears in a linking expression, one can omit its
index, as depicted in line~\ref{lst:node-example-l1} of
Fig.~\ref{fig:node-example}.  In our language---unlike in Click---one
can connect outputs to configuration arguments as well, which enables
the flexible dynamic configuration of nodes.  This uses the same
syntax as the output--input linking.  To distinguish between inputs
and configuration arguments, a hyphen should be prepended to the
one-based index of the configuration arguments (see
lines~\ref{lst:tsg-lh}--\ref{lst:tsg-li} of Listing~\ref{lst:tsg} for
an example).  We created a special linking operator, the \viewLink{},
that can be used when connecting the node itself to a View, as used in
line~\ref{lst:node-example-l2} of our example\footnote{In this case
  the \texttsg{Graph} node does not have any outputs but is
  responsible for displaying graph visualization on its own.}.
Terminating an expression (a node declaration or a series of linkings)
should always be done with a semicolon.

\subsection{Execution Framework for TSGs}

To bring the TSG concept closer to implementation, we designed an
execution framework capable of interpreting TSGs, executing them and
providing intelligent navigation among the (possibly many) outputs of
the tools used in the TSG.

\subsubsection{Interpretation}

Since a \gls{tsg} is only a formal description, the framework provides
a parser to interpret the graph.  Two methods are available to
accomplish this. The first option is building the graph by
interpreting its definition written in our description language.
Using this method, graphs can be saved and later reused and shared.
The parser collects the nodes, their configuration arguments and the
links between the nodes.  Objects are created for nodes and edges. The
framework assigns names to node and edge objects that uniquely
identifies them. The second option is creating the graph incrementally
at run time using a drag and drop method.  This method has the benefit
that the operator can monitor the output of the used tools and adjust
her methodology to the results she has acquired until that point
(which is fully in line with current practice). The parser creates the
objects for the nodes and interconnections, and creates a description
for these using the language.  Run-time modification of the graph is
provided by this functionality as well.  These two methods provide
flexibility while retaining the benefit of being able to store
troubleshooting scenarios, situations.

In order to help writing a \gls{tsg} definition file, the framework
provides syntax highlighting to differentiate semantic units.  Using
the nodes' self-documentation capabilities, the framework is able to
provide on-the-fly node documentation.

\subsubsection{Execution}

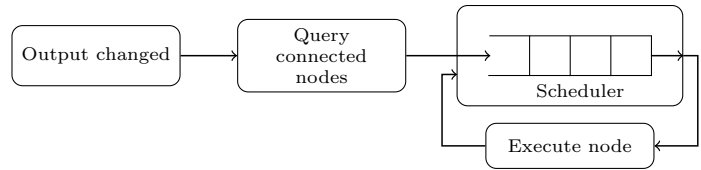
\begin{figure}
\centering
\begin{tikzpicture}[node distance=2cm, font=\sffamily]

  \node [rectangle, rounded corners, minimum width=2cm, minimum height=.8cm, text centered, text width=2cm, font=\scriptsize, draw=black] (change) {Output changed};

  \node [rectangle, rounded corners, minimum width=2.2cm, minimum height=.8cm, text centered, text width=2cm, font=\scriptsize, draw=black, right of=change, xshift=1cm] (query) {Query \mbox{connected} nodes};

  \node [rectangle, rounded corners, minimum width=3cm, minimum height=1cm, text centered, text width=2cm, font=\scriptsize, draw=black, right of=query, xshift=1.3cm] (scheduler) {{\parbox[b][1.1cm]{.9cm}{Scheduler}}};

    \draw [->, line width=.2mm] (change) -- (query);
      
  \node (queue) [three sided, rotate=180, minimum width=.5cm, minimum height=.5cm, left of=query, xshift=-.5cm] {};
  \node (queue1) [three sided, rotate=180, minimum width=.5cm, minimum height=.5cm, left of=queue, xshift=1.46cm]  {};
  \node (queue2) [three sided, rotate=180, minimum width=.5cm, minimum height=.5cm, left of=queue1, xshift=1.46cm]  {};
  \node (queue3) [three sided, rotate=180, minimum width=.5cm, minimum height=.5cm, left of=queue2, xshift=1.46cm]  {};

  \node [rectangle, rounded corners, minimum width=2cm, minimum height=.6cm, text centered, text width=2cm, font=\scriptsize, draw=black, below of=scheduler, yshift=.8cm] (execute) {Execute node};

  \draw [->, line width=.2mm] (query) -- (queue);
  \draw [->, line width=.2mm] (queue3) -- (scheduler);

  \draw [->, line width=.2mm] (scheduler) -- (8, 0) -- (8, -1.2) -- (execute);
  \draw [->, line width=.2mm] (execute) -- (4.6, -1.2) -- (4.6, -.25) -- (4.8, -.25);
  
\end{tikzpicture}
\caption{Scheduling of node execution.}
\label{fig:scheduler}
\end{figure}

The \gls{tsg} concept describes how to connect tools to each other but
it does not deal with the problem of when and how a node's life cycle
is managed and how a node is notified when its inputs are updated.
After interpretation, the framework creates placeholders for nodes
which they can then use for displaying graphical data.  Once \gls{tsg}
execution is started, the framework is responsible for handling node
execution, as depicted in Fig.~\ref{fig:scheduler}.  The framework
creates a scheduler that handles a queue for registering node output
changes.  Each time an output has changed, the framework queries which
nodes have that as their input.  These destination nodes are then
inserted to the end of the queue.  Parallel to this, a simple
scheduling mechanism is running that always takes the first item from
the beginning of the queue and sends a signal to that node to enter
into the execution stage.  When the call returns, the scheduler moves
to the next node in line and so on.

\subsubsection{Navigation Options}

The benefit of handling interconnected troubleshooting tools as a
graph is that it creates a natural order in the troubleshooting
process.  In order to better manage the complex information set
contained in the graph, the framework provides different apparatuses
to aid observing the execution state and navigating through the graph.
A special node class, a \texttsg{View}, is available that can collect
nodes or their outputs and group those to be displayed together.  When
a name appears without a class assignment, the interpreter assumes
that it is a \texttsg{View}.  Layout of such groupings can be
explicitly specified or left to the framework to create an automatic
one. The framework provides methods for graph traversal by jumping
through successive nodes and edges and offers a semantically ordered
grouping for nodes and edges contained in the currently displayed
\gls{tsg}.  It also provides a visualization method that creates a
graphical representation of the \gls{tsg}.

\subsection{Recommendation System and Knowledge Sharing}
  
The framework provides a recommendation system that is able to suggest
new nodes for the operator, based on her current setup, by searching
for similarities in a TSG repository. Operators can upload their
existing TSGs to this repository hereby promoting knowledge sharing.

\section{An Everyday Example}
\label{sec:example}

For showcasing our framework, we present here an everyday example that
(or very similar cases), almost surely, appears in the practice of
every operator.  We set up a Mininet emulated network to create
failures and compare manual troubleshooting with our \gls{tsg}-based
methodology.  In our network, hosts are connected to servers via
OpenFlow switches and ordinary routers---emulated as hosts having the ability
to route between their interfaces.  The switches act as simple layer 2
devices, we use only static routes and \iptables{} is applied for
creating firewall rules.  \ssh{} access is set up for every device in
the network.

We inject various network related errors to cause the most basic
symptom of a network issue: a host on the network is not able to
connect to a server.  Emulated errors encompass hardware
failures---link and port failures---, configuration errors---faulty
configuration of hosts, misconfigured routes and firewall rules on
routers---and application level errors---misconfigured or unresponsive
applications. Our task is to identify and correct the error: first in
line with current practice and then, using our framework.

\subsection{Current Practice}

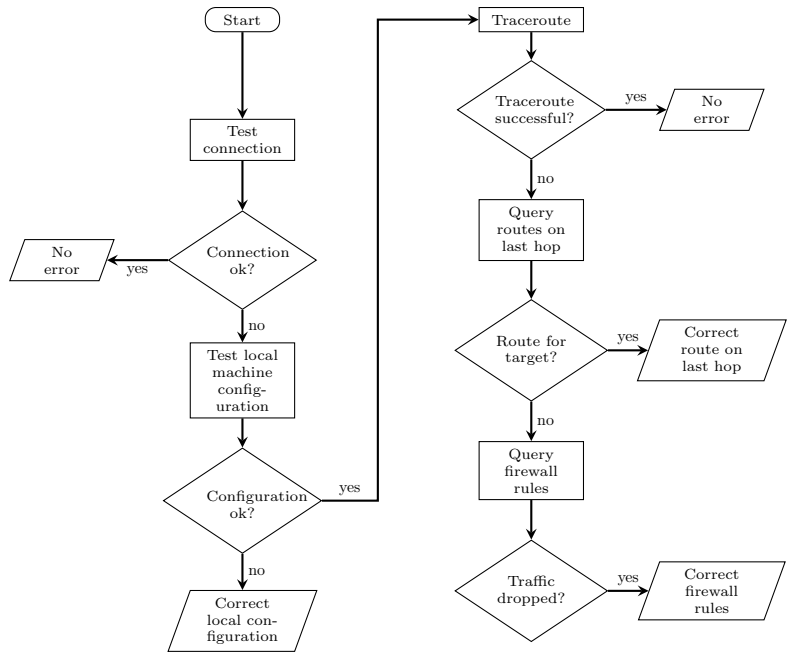
\begin{figure}
  \centering
\begin{tikzpicture}[node distance=2cm]

\begin{scope}[every node/.style={scale=.8}]
\node (start) [startstop] {Start};
\node (t_conn) [process, below of=start] {Test connection};
\node (d_conn) [decision, below of=t_conn, xshift=0cm, aspect=1.5] {Connection ok?};
\node (no_err) [io_small, left of=d_conn, yshift=-0.0, xshift=-1.0cm] {No error};
\node (t_conf) [process, below of=d_conn, xshift=0cm] {Test local machine configuration};
\node (d_conf) [decision, below of=t_conf, xshift=0cm, aspect=1.5] {\hspace{1cm}\mbox{Configuration} ok?};
\node (corr_conf) [io, below of=d_conf, yshift=0.0cm, xshift=0.0cm] {Correct local configuration};
\node (trace) [process, right of=start, xshift=2.8cm] {Traceroute};
\node (d_trace) [decision, below of=trace, yshift=0.5cm, aspect=1.5] {Traceroute \mbox{successful}?};
\node (no_err2) [io_small, right of=d_trace, yshift=-0.0, xshift=1.0cm] {No error};
\node (route) [process, below of=d_trace, yshift=0.0cm] {Query routes on last hop};
\node (d_route) [decision, below of=route, xshift=0cm, aspect=1.5] {Route for target?};
\node (corr_route) [io, right of=d_route, yshift=-0.0cm, xshift=1.0cm] {Correct route on last hop};
\node (fw) [process, below of=d_route, xshift=0cm] {Query firewall rules};
\node (d_fw) [decision, below of=fw, xshift=0cm, aspect=1.5] {Traffic dropped?};
\node (corr_fw) [io, right of=d_fw, xshift=1.0cm] {Correct firewall rules};
\end{scope}

\begin{scope}[every node/.style={scale=.6}]
\draw [arrow] (start) -- (t_conn);
\draw [arrow] (t_conn) -- (d_conn);
\draw [arrow] (d_conn) -- node[anchor=north] {yes} (no_err);
\draw [arrow] (d_conn) -- node[anchor=west] {no} (t_conf);
\draw [arrow] (t_conf) -- (d_conf);
\draw [arrow] (d_conf) --  node[anchor=west] {no} (corr_conf);
\draw [arrow] (d_conf) -- node[anchor=south] {yes} (1.8, -6.4) |- (trace);
\draw [arrow] (trace) -- (d_trace);
\draw [arrow] (d_trace) -- node[anchor=south] {yes} (no_err2);
\draw [arrow] (d_trace) -- node[anchor=west] {no} (route);
\draw [arrow] (route) -- (d_route);
\draw [arrow] (d_route) -- node[anchor=south] {yes} (corr_route);
\draw [arrow] (d_route) -- node[anchor=west] {no} (fw);
\draw [arrow] (fw) -- (d_fw);
\draw [arrow] (d_fw) -- node[anchor=south] {yes} (corr_fw);
\end{scope}

\end{tikzpicture}
\caption{Flow chart for troubleshooting the exemplary scenario.}
\label{fig:flow_chart}
\end{figure}

By applying manual troubleshooting using multiple shells to connect to
different devices and running different troubleshooting tools, we made
the following observations.  We had to repeatedly log into devices
when starting a new tool and for every new tool, a new terminal had to
be opened.  Processes running in currently open terminals could have
been terminated in order to minimize the overpopulation of our
environment but than we would have lost information.  We had to
analyze the data provided by the tools ourselves and it took a
considerable time when we were dealing with great amounts of data.
Finally, we noticed that the list of applicable troubleshooting tools
could be narrowed down to a minimum and then it did not matter what
the root cause was, we ended up using the same tools in the same
sequence to find the problem.  Thus we came up with a flow chart for
this troubleshooting scenario (see Fig.~\ref{fig:flow_chart}).  One
major problem with the multiple shell approach is that there is no
clear way that the process---the steps to be taken---can be recorded
and later reused.

To test connectivity we used \texttsg{ping}. For checking the local
host's configuration we applied the \texttsg{host}, \texttsg{ifconfig}
and \texttsg{arp} tools.  Forwarding rules were queried by
\texttsg{route}, and firewall rules were retrieved using
\texttsg{iptables}.  The steps shown in Fig.~\ref{fig:flow_chart}
could have been grouped together into a single shell script file for
automation purposes but we found that method less clear and portable.

\subsection{Using \glspl{tsg}}

Now we show that the above process can be easily mapped\footnote{With
  an appropriate node repository only the connections need to be
  correctly specified among nodes.} to a \gls{tsg}, as most of the
nodes are already identified by the processes and decisions on the
flow chart (Fig.~\ref{fig:flow_chart}).  We use extended wrapper nodes
for the tools, so---on top of their basic functionality---we assume that
\texttsg{Host}, \texttsg{Ifconfig} and \texttsg{Arp} nodes provide
means to exclude specified interfaces from their outputs, the
\texttsg{Traceroute} node is able to relay the last hop until which
the traffic is traceable and the \texttsg{Route} and
\texttsg{Iptables} nodes are capable of displaying only those rules
that apply to certain IP addresses.

For brevity, Listing~\ref{lst:tsg} shows only\footnote{See the
  complete TSG in Appendix~\ref{app:fulltsg}.} the instructions used
for creating a \gls{tsg} that performs a connection test and if that
is unsuccessful, checks the local machine's network configuration.
This exemplary \gls{tsg} is not complete: routes and firewall rules
are not checked.  Instructions for creating nodes for those tests,
however, would be written using the same philosophy shown here.
\texttsg{Decision} nodes in the \gls{tsg} take the same role as the
decisions shown on the flow chart of Fig.~\ref{fig:flow_chart}, while
the \texttsg{Decision-summary} node interprets decisions in the
\gls{tsg} by signaling ``error'' codes with visual aids for the
operator. (Decision nodes are only briefly discussed here,
see Section~\ref{sec:branches} for more details.)

\begin{lstlisting}[float, floatplacement=H, language=C++, showstringspaces=false,
basicstyle=\ttfamily\footnotesize, columns=fullflexible, numbers=left,
xleftmargin=2em, caption=Formal description of the \gls{tsg}., label=lst:tsg, escapeinside={(*@}{@*)}]
ping :: Ping(localhost, <address of the server>);(*@\label{lst:tsg-la}@*)
ifconfig :: Ifconfig(localhost);(*@\label{lst:tsg-lb}@*)
arp :: Arp(localhost, nil, -n);(*@\label{lst:tsg-lc}@*)

ping-decision :: Decision(..., string-match,
                                   ..., ttl);
ifc-decision :: Decision(...,
                                 ifconfig-check-interfaces,
                                 ..., lo);
arp-decision :: Decision(..., (lambda (x)
                                           (> (length x) 0)),
                                  ...);

ping -> ping-decision;(*@\label{lst:tsg-ld}@*)
ifconfig -> ifc-decision;
arp -> arp-decision;(*@\label{lst:tsg-le}@*)

ping-decision[1] -> ifconfig;(*@\label{lst:tsg-lf}@*)
ifc-decision -> Function(ifconfig-get-interfaces,(*@\label{lst:tsg-lg}@*) 
                                 'input-0)[0, 0](*@\label{lst:tsg-lh}@*)
->  [0, -2]arp;(*@\label{lst:tsg-li}@*)

ping-decision[2] -> ds :: Decision-summary();(*@\label{lst:tsg-lj}@*)
ifc-decision[2] -> [1]ds;
arp-decision[2] -> [2]ds;(*@\label{lst:tsg-lk}@*)
\end{lstlisting}

\begin{figure}
  \centering
  \includegraphics[width=3.7in]{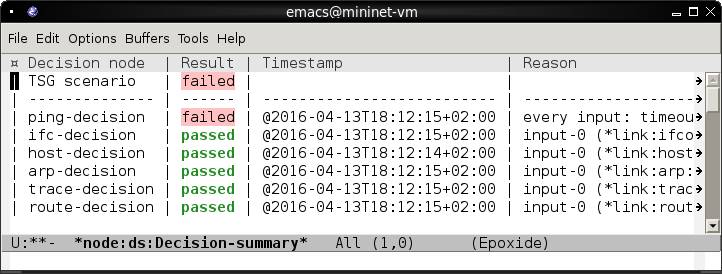}
  \caption{Summarizing a troubleshooting scenario.}
  \label{fig:summary}
\end{figure}

In order to perform a connection test between the current host and a
server, we use a wrapper node for the \texttsg{ping} tool in
line~\ref{lst:tsg-la}.  Lines~\ref{lst:tsg-lb}--\ref{lst:tsg-lc}
create tests for checking the local host's configuration using the
\texttsg{ifconfig} and \texttsg{arp} tools.  In order to automatically
evaluate these, we defined three Decision nodes.
\texttsg{ping-decision} checks whether the \texttsg{ping} was
successful.  \texttsg{ifc-decision} uses a custom function to validate
the interface configuration returned by the \texttsg{ifconfig}
node---the \emph{loopback} interface is ignored by the test for
obvious reasons.  The \texttsg{arp-decision} node performs a simple
check to test whether there are entries in the \gls{arp} cache.  These
Decision nodes are then connected with their respective wrapper
nodes in lines~\mbox{\ref{lst:tsg-ld}--\ref{lst:tsg-le}}.  In order to
check the local host's configuration only when the connection test was
unsuccessful, we need to connect the \texttsg{ifconfig} node to the
negative output of the \texttsg{ping-decision}
node---line~\ref{lst:tsg-lf} implements that.  If there are interfaces
on the host that are correctly configured, we need to check whether
the host can register the layer 2 addresses from its network.  We
defined a \texttsg{Function} node in line~\ref{lst:tsg-lg} in order to
retrieve the interface names from the output of the
\texttsg{ifc-decision} and fed these to the our \texttsg{Arp} node.
Finally, we defined a \texttsg{Decision-summary} node in
lines~\mbox{\ref{lst:tsg-lj}--\ref{lst:tsg-lk}} to display information
collected from every Decision node in a summarizing table.  A possible
output (generated with our prototype) of the Decision-summary node is
shown in Fig.~\ref{fig:summary}. The node gives the results of the
individual decisions in the \gls{tsg} as well as an overall evaluation
of the current troubleshooting scenario.

This simple example attests that by using \glspl{tsg}, we can achieve
a state of automation where the parametrization of troubleshooting
tools adapts to the current network situation and issue.  When
executing this \gls{tsg}, the operator can recognize failure modes at
a glance by looking at the error codes, or can further delve into
details by navigating through the outputs of nodes.  Using the
navigation options provided by the framework, the operator can walk
through the troubleshooting process in an orderly fashion.  Results
are going to be displayed according to the workflow, she laid out when
she had designed the flow chart for locating the issue.  Even more,
the operator has the means to hide irrelevant data by grouping nodes
and edges together with Views, or at runtime select only those, she
deems relevant to the situation at hand.

By offering proper formalization, \glspl{tsg} can be reused in similar
scenarios.  In some cases, the whole graph can be reused with only
slight adjustments to the node configuration arguments to adapt to
other network conditions.  In other cases, subgraphs of the original
\gls{tsg} can be reused for testing different scenarios.  Besides
re-usability, \glspl{tsg} can act as a technique to collect
troubleshooting know-hows.  Once a network problem is uncovered using
a \gls{tsg}, it automatically becomes a guideline for discovering
future similar issues.  By collecting a library of these, operators
can greatly decrease problem solution times and the efficiency of
knowledge transfer to new operators can also increase.  If \glspl{tsg}
are not only collected within a closed environment---i.e.~within a
company---but are shared with a greater audience, they can prove to be
beneficial for the whole networking community.  If a wide \gls{tsg}
library is paired with problem descriptions and solutions, new nodes
and test cases can be recommended to an operator, based on previous
cases described by \glspl{tsg} in the library.

\section{Prototype}
\label{sec:implementation}

For proofing the concept of our framework, we created a prototype
implementation named \epoxide{} using GNU \emacs{} as a platform.
\emacs{} is an extensible, customizable text editor, its central
concept, the buffer, is responsible for holding file contents,
subprocess outputs, providing configuration interfaces, etc.  By its
extensible nature, \emacs{} offered a particularly good platform to
build \epoxide{} upon.  \emacs{} supports, via its advanced text
manipulation and documentation functions, writing \gls{tsg}
definitions that we store in \tsgFile{} configuration files.  We
implemented our execution framework and wrapper nodes using \emacs{}'s
own Lisp dialect, \emacslisp{}.  Nodes and their outputs are assigned
to \emacs{} buffers for observability.  Node interface and connection
information is stored in buffer local variables.  We note that while
\emacs{} proved to be a good fit for our notions, the concept
described in Section~\ref{sec:principles} can be implemented on other
platforms as well.

\subsection{Framework Functions}

Opening a \tsgFile{} file in \emacs{} automatically loads \epoxide{}
and interprets the \gls{tsg} stored in it.  After assigning node
objects and outputs to buffers, nodes can write in these buffers
directly during their execution stage.  In our prototype
implementation, nodes use other nodes' output buffers directly as
their inputs.  Thus \gls{tsg} edges do not have independent objects.

\emacs{}'s \emph{\eldoc{}} package uses the nodes' self-documentation
property to provide hints on node interfaces during \gls{tsg}
definition.  When \emacs{}'s \emph{\autocomplete{}} package is
installed, it can provide intelligent code completion for setting up
nodes.

Once a \gls{tsg} is running, there are two ways to modify edges and
node configuration arguments.  The first method is runtime \gls{tsg}
reconfiguration, which is provided via an \emacs{} widget based
interface.  It allows the addition of \gls{tsg} edges and modification
of static configuration arguments. Modifications are committed to the
\tsgFile{} file as well.  The second method a \glspl{tsg} can be
modified is by way of editing the \tsgFile{} file and reevaluating it.
This method has the downside that buffer contents prior to the
modification get lost.

Once a \gls{tsg} is processed, execution is started.  The framework
stores events in a simple \emacslisp{} list and utilizes the
\mbox{\emph{timerfunctions}} package for scheduling node execution.
Basic navigability among the created buffers is provided by \emacs{}
key bindings.  \epoxide{} supplies the apparatus to move from one node
buffer to its output buffers or to the next node's buffer (in forward
or backward direction).  To traverse a \gls{tsg}, a visualization can
be used also, supported by the \emacs{} \emph{\cogre{}} package.  When
the \emph{\graphviz{}} external software is installed, the displayed
graph can be drawn using an automatic layout for better visual
clarity.  Semantic grouping of the created buffers is provided using
the \emph{\ibuffer{}} package: a dynamic list is displayed that
aggregates buffers based on their types and roles in the current
context.

The current implementation of \epoxide{} provides a module for
collecting \gls{tsg} related data and supplying node recommendations.
We created the instrumentation for basic case-based reasoning where
currently available \glspl{tsg} are considered as descriptions of
previous troubleshooting cases.  These \glspl{tsg} are indexed and
their data---together with information from the current \gls{tsg}---is
passed to a recommender.  The recommender then can suggest nodes based
on these pieces of data.  We created an interface in the framework to
which recommenders, developed by third parties, can connect as well.
By now, we have implemented the most basic recommender that suggests
the most popular nodes and displays them using \emacs{}'s
\emph{\ido{}} package.  Most popular nodes are computed by counting
every node in all previously written \glspl{tsg} and ordering them in
the descending order of their cumulative count.  Nodes that are
already used in the current \gls{tsg}, are excluded from the
suggestion list.

\subsection{Branching Nodes}
\label{sec:branches}

When creating the apparatus that enables conditional branching in
\epoxide{}, the most basic expectation was to
\begin{enumerate*}[label=\emph{(\roman*)}]
\item \label{enum:1} provide functionality to analyze and evaluate the
  output of any node and, based on the result,
\item \label{enum:2} create branches in a \gls{tsg}, the same way a
  decision would in a flow chart.  Additionally, we wanted to have the
  ability to
\item \label{enum:4} select among different inputs or to use a
  combination of them.  This criterion was inspired by how nodes work
  in a Bayesian network: they receive the results of lower level nodes
  and calculate their outputs based on that.
\end{enumerate*}
To satisfy these criteria, we implemented a single \emph{Decision
  node}. Since nodes can be added to \epoxide{} in a modular fashion,
this is only one possible implementation that satisfies our initial
criteria.  Operators are free to add their own version. A
\emph{Decision-summary node} was developed in conjunction, for
summarizing the outputs of such nodes.

\begin{figure}[!t]
\centering
\begin{tikzpicture}[node distance=2cm]

\node (tool1) [process] {Tool 1};
\node (dots1) [rectangle, below of=tool1, yshift=1cm] {\dots};
\node (tooln) [process, below of=dots1, yshift=1cm] {Tool n};

\node [rectangle, minimum width=6cm, minimum height=3.5cm, text centered, text width=2cm, font=\scriptsize, draw=black, right of=dots1, xshift=2.75cm, yshift=-.25cm] (decision-node) {};

\node (preproc1) [rectangle, minimum width=.2cm, minimum height=.2cm, text centered, text width=1.7cm, font=\scriptsize, draw=black, right of=tool1, xshift=1.5cm] {Verify input 0};
\node (dots2) [rectangle, right of=dots1, xshift=1.5cm] {\dots};
\node (preprocn) [rectangle, minimum width=.2cm, minimum height=.2cm, text centered, text width=1.7cm, font=\scriptsize, draw=black, right of=tooln, xshift=1.5cm] {Verify input n};

\node (selection) [rectangle, xshift=2cm, yshift=2cm, minimum width=2.45cm, rotate=-90, minimum height=.7cm, text width=2.1cm, text centered, font=\footnotesize, draw=black, right of=dots2] {Combine or select};

\node (pout) [rectangle, right of=preproc1, xshift=1.6cm, yshift=.15cm] {\scalebox{.5}{Positive output}};
\node (pout) [rectangle, right of=dots2, xshift=1.58cm, yshift=.15cm] {\scalebox{.5}{Negative output}};
\node (pout) [rectangle, right of=preprocn, xshift=1.65cm, yshift=.15cm] {\scalebox{.5}{Status output}};

\node (pout) [rectangle, below of=preprocn, xshift=1cm, yshift=1.3cm] {Decision node};

\draw (6.3, -1) circle (.15cm);

\draw [arrow] (tool1) -- (1.75, 0);
\draw [arrow] (tooln) -- (1.75, -2);
\draw [arrow] (1.75, 0) -- (preproc1);
\draw [arrow] (1.75, -2) -- (preprocn);
\draw [arrow] (preproc1) -- (5.1, 0);
\draw [arrow] (preprocn) -- (5.1, -2);

\draw [arrow] (5.9, -1) -- (6.15, -1);

\draw [arrow] (6.3, -.85) -- (6.3, 0) --  (7.75, 0);
\draw [arrow] (6.45, -1) -- (7.75, -1);
\draw [arrow] (6.3, -1.15) -- (6.3, -2) -- (7.75, -2);

\draw [arrow] (7.75, 0) -- (8.75, 0);
\draw [arrow] (7.75, -1) -- (8.75, -1);
\draw [arrow] (7.75, -2) -- (8.75, -2);

\end{tikzpicture}
\caption{Schematics of a Decision node.}
\label{fig:decision}
\end{figure}
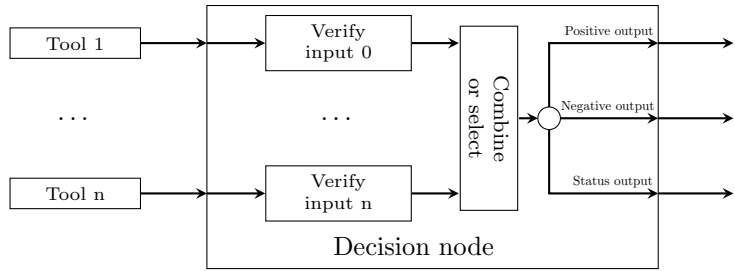

Fig.~\ref{fig:decision} shows the schematics of our Decision node.
The node can attach to any wrapper node\footnote{The node is also
  prepared to handle the asynchronism and the different output formats
  of the wrapper nodes.} and incoming data is first verified (as
per~\ref{enum:1}): it is determined whether it complies with certain
criteria.  The node can use any function for verification that has at
least one argument (the input) and can return false, when verification
failed, or any string otherwise.  When verification of the inputs is
finished, further processing can be done using a second stage
function, in accordance with~\ref{enum:4}.  An operator can use a
function to select an input with, for example, the \emph{or} function:
the first of the inputs that passed verification is going to be the
result of this stage.  Other functions can implement more complex
processing, like in the aforementioned Bayesian network example where
the result would be a probability value.  Such second stage functions
should return a string in case of a positive decision or false
otherwise.  We defined two node outputs to relay these return values
(as per~\ref{enum:2}).  The first displays information in the positive
case, the second in the negative.  We implemented an additional output
for interoperating with Decision-summary nodes.

\subsection{Implementing Nodes}

The modular architecture of \epoxide{} makes it possible for anyone to
implement new nodes\footnote{The list of nodes currently implemented
  in \epoxide{} is shown in Appendix~\ref{app:listofnodes}.}.  A node
developer---who implements a node---has to create separate node
definition files that contain \emacslisp{} code to provide
self-documentation for nodes and implement the three life cycle stages
via functions.  For proper interaction with nodes, the node definition
files and these functions should comply with a fixed naming scheme.
Node self-documentation functions should provide information about
node interfaces and could also implement validation functions to check
the compliance of arguments with certain criteria.  The functions
implementing the separate node life cycle stages are called by the
framework on specific occasions.  The initialization function is
evoked after the buffers belonging to the node are set up.  The
execution function is called every time a change occurs on any of the
node's inputs.  Finally, the termination function is invoked before
the framework closes the buffers associated with the node.  These
functions can draw on common node functions provided by the framework.
\epoxide{} implements functions for setting up node buffers with basic
data, reading inputs and configuration arguments, writing outputs and
handling remote access using \emacs{}'s \emph{\glsentryfull{tramp}}
package.

\section{Case studies}
\label{sec:case}
The basic example shown in Section~\ref{sec:example} demonstrated
integration of troubleshooting tools originally developed for
traditional networks.  Although these can also be used to troubleshoot
\glspl{sdn}, tools created specifically for these networks can
leverage the benefits of centralized network control.

\subsection{Troubleshooting in \glspl{sdn}}
\label{sec:applicability-for-sdn}

We implemented nodes to interact with POX, Floodlight and OpenDaylight
controllers and \gls{ovs} switches.  These nodes are able to collect
\glspl{dpid}, flow statistics and topology information.  Additionally,
we created processing nodes for filtering flow statistics on a flow
space\footnote{Those flow table entries make up a flow space that
  match given source and destination point pairs.} basis and for
visualizing topology information.  In order to aid controller
debugging, we wrapped the \gls{gdb} tool as well. To support other
controller platforms, a general node for accessing REST APIs was
implemented, as well as other nodes that are able to filter specific
parts of JSON-formatted textual data.

\begin{figure}
  \centering
  \subfloat{
    \resizebox{9cm}{5.5cm}{
      \begin{tikzpicture}[node distance=4cm, font=\sffamily]
        
        \node [shape=nodea] (clock) {Clock};

        \node [shape=nodeb, right of=clock] (dpid) {Dpids};

        \draw [->] (clock.outputa) -- (dpid.inputa);

        \node [shape=function, right of=dpid] (function) {Function};
        
        \draw [->] (dpid.outputa) -- (function.inputa);

        \node [shape=nodeb, below of=clock, yshift=1.5cm] (fs1) {Flow-stat};

        \draw [->] (function.outputa) -- (10, .5) -- (10, -1.35) -- (-1.75, -1.35) -- (-1.75, -2) -- (fs1.inputa);

        \node [shape=nodeb, below of=fs1, yshift=1.5cm] (fs2) {Flow-stat};

        \draw [->] (function.outputb) -- (9.75, 0) -- (9.75, -1.15) -- (-2, -1.15) -- (-2, -4.5) -- (fs2.inputa);

        \node [shape=nodeb, right of=fs1] (filter1) {Flow-space-filter};

        \draw [->] (fs1.outputa) -- (filter1.inputa);

        \node [shape=nodeb, right of=fs2] (filter2) {Flow-space-filter};

        \draw [->] (fs2.outputa) -- (filter2.inputa);

        \node (dots) [rectangle, below of=filter2, yshift=2.5cm] {\vdots};

        \node [shape=table, right of=filter2, yshift=1.5cm] (table) {Table-view};

        \draw [->] (filter1.outputa) -- (6, -2) -- (6, -3) -- (table.inputa);

        \draw [->] (filter2.outputa) -- (6, -4.5) -- (6, -3.5) -- (table.inputb);
                
      \end{tikzpicture}}}
  \hfill
  \subfloat{
  \includegraphics[width=3.7in]{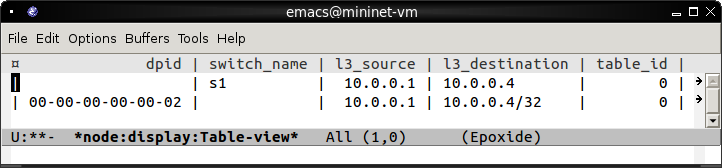}}
  \caption{An \gls{sdn} example: querying flow statistics with a \gls{tsg} (top) and displaying them in a table using a \texttsg{Table-view} node (bottom).}
  \label{fig:sdn-case-study}
\end{figure}
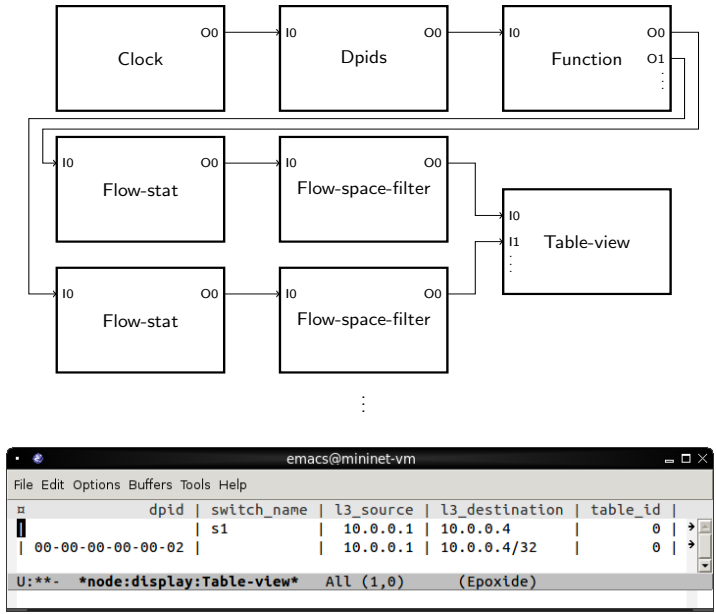

To illustrate an \gls{sdn} use-case, imagine the following scenario: a
network consisting of \gls{sdn} switches is given and an operator
wants to monitor a certain traffic flow in the network.  She can set
up a \gls{tsg} to solve this problem by first defining a node querying
\glspl{dpid} from the controller on a timely basis, as depicted in the
top part of Fig.~\ref{fig:sdn-case-study}.  A \texttsg{Function} node
can then separate the list of available \glspl{dpid} and pass those to
further nodes that query flow statistics from their assigned switches.
With the addition of flow space filtering nodes, the operator can
select only the flow under scrutiny and display the results in a table
format, as depicted by the bottom part of
Fig.~\ref{fig:sdn-case-study}.  She can investigate further by adding
nodes that query flow statistics from the controller and from the
switches themselves, and compare the results with \texttsg{Decision}
nodes: this could reveal synchronization issues between the devices or
misdirection of the traffic caused by a software failure.  By
connecting \texttsg{Gdb} nodes with the \texttsg{Decision} nodes, the
operator can remotely connect to software switches or to the
controller for an in-depth analyzation of the problem.\footnote{A live
  action demo on a similar, albeit simpler, case can be watched
  at~\cite{levai2015epoxide}:
  \linebreak\url{https://www.youtube.com/watch?v=HsiGFR0QirE}}

\subsection{Troubleshooting in Service Function Chaining}

One of the main goals of the UNIFY
project\footnote{\url{https://www.fp7-unify.eu}} was to design a
Service Function Chaining control plane architecture and implement a
proof-of-concept prototype.  Additionally, the project also introduces
the concept of Service-Provider DevOps to combine the developer and
operational workflows in carrier grade environments.  DevOps results
in faster deployment cycle of novel networking services.  Instead of
designing complex services as a whole, these services are assembled
from atomic \emph{network functions}.  However, fast deployment cycles
require faster testing phases and troubleshooting in the operational
environment.  Even when a new service is created by re-using
components of previous ones, it is still going to be different enough
from earlier scenarios to implicate new troubleshooting challenges.
In these cases, previous knowledge is not always directly applicable.
By providing an integrated platform for running troubleshooting tools
and an apparatus to automatize their execution, our tool makes the
formation of new troubleshooting scenarios easier, thus enabling
quicker service deployment.

\epoxide{} has a central role in the multipurpose demonstrator
showcasing major results of the project~\cite{D43}.  Using its
dedicated and general purpose wrapper nodes, it orchestrates multiple
components in a semi-automatic troubleshooting scenario.  \epoxide{}
needs to reveal a configuration error resulting in erroneous imbalance
of the traffic loads of OpenFlow switches instantiated as network
functions.  The novel components, which \epoxide{} interacts with,
include a flexible messaging bus (Double Decker), a tool that
calculates aggregated performance metrics derived from data queried
from hierarchical time series databases [Recursive Query Engine
(RQE)], and a test packet generator for pinpointing errors in OpenFlow
switches (AutoTPG).\footnote{Detailed description of these tools and
  the demonstrator can be found in~\cite{D43}.}

\begin{figure}
\centering
\begin{tikzpicture}[node distance=2cm, font=\sffamily]

  \node (s) [coordinate] {};
  \node (e) [coordinate, right of=s, xshift=7.5cm] {};

  \draw [-, line width=.2mm] (s) -- (e);

  \node [rectangle, minimum width=1cm, minimum height=2cm, text centered,
text width=1cm, font=\scriptsize, draw=black, below of=s, xshift=.9cm,
yshift=.3cm] (epoxide) {\epoxide{}};

  \node [rectangle, minimum width=1cm, minimum height=.8cm, text centered,
text width=1cm, font=\scriptsize, draw=black, right of=epoxide,
xshift=.2cm, yshift=.6cm] (rqe) {RQE};

  \node [rectangle, minimum width=1cm, minimum height=.8cm, text centered,
text width=1cm, font=\scriptsize, draw=black, below of=rqe, yshift=.8cm]
(autotpg) {AutoTPG};

  \node [rectangle, minimum width=1.5cm, minimum height=.8cm, text
centered, text width=1.5cm, font=\scriptsize, draw=black, right of=rqe,
xshift=.3cm] (tsdb) {Time Series Database};

  \node [rectangle, minimum width=1.5cm, minimum height=.8cm, text
centered, text width=1.5cm, font=\scriptsize, draw=black, right of=autotpg,
xshift=3.4cm, yshift=-.1cm] (nf2) {};

  \node [rectangle, minimum width=1.5cm, minimum height=.8cm, text
centered, text width=1.5cm, font=\scriptsize, draw=black, right of=autotpg,
xshift=3.35cm, yshift=-.05cm, fill=white] (nf1) {};

  \node [rectangle, minimum width=1.5cm, minimum height=.8cm, text
centered, text width=1.5cm, font=\scriptsize, draw=black, right of=autotpg,
xshift=3.3cm, fill=white] (nf) {Network Functions};

  \node [rectangle, minimum width=1.5cm, minimum height=.8cm, text
centered, text width=1.5cm, font=\scriptsize, draw=black, right of=tsdb,
xshift=1cm] (tmc) {Traffic Monitoring Component};

  \draw [-, line width=.2mm] (1.51, -1.1) -- (rqe);
  \draw [-, line width=.2mm] (rqe) -- (tsdb);
  \draw [-, line width=.2mm] (1.51, -2.3) -- (autotpg);
  \draw [-, line width=.2mm] (tmc) -- (nf);

  \node [rectangle, minimum width=5cm, minimum height=.1cm, text centered,
text width=5cm, font=\scriptsize, right of=s, xshift=5.85cm, yshift=.2cm] (mb)
{Messaging bus (Double Decker)};

  \node [rectangle, minimum width=1cm, minimum height=2cm, text centered,
text width=1cm, font=\scriptsize, right of=epoxide, xshift=-.9cm,
yshift=.7cm] (restapi1) {\scalebox{.6}{REST API}};

    \node [rectangle, minimum width=1cm, minimum height=2cm, text centered,
text width=1cm, font=\scriptsize, right of=epoxide, xshift=-.9cm,
yshift=-.5cm] (restapi1) {\scalebox{.6}{REST API}};

  \draw [-, line width=.2mm] (epoxide) -- (.9, 0);
  \draw [fill=black] (.9, 0) circle (.02cm);

  \draw [-, line width=.2mm] (tmc) -- (8.4, 0);
  \draw [fill=black] (8.4, 0) circle (.02cm);

  \draw [<-, line width=.2mm] (tsdb) -- (tmc);
  \draw [-, line width=.2mm] (autotpg) -- (nf);

  \draw [dashed, ->, line width=.2mm] (8.3, -.59) -- (8.3, -.1) -- (1, -.1) -- (1, -.7);
  \draw (5, -.25) circle (.1cm) node {\scalebox{.5}{1}};

  \draw [dashed, ->, line width=.2mm] (1.52, -.8) -- (2.48, -.8);
  \draw (2, -.65) circle (.1cm) node {\scalebox{.5}{2}};

  \draw [dashed, ->, line width=.2mm] (2.48, -1.3) -- (1.52, -1.3);
  \draw (2, -1.45) circle (.1cm) node {\scalebox{.5}{5}};

  \draw [dashed, ->, line width=.2mm] (3.72, -.8) -- (4.54, -.8);
  \draw (4.1, -.65) circle (.1cm) node {\scalebox{.5}{3}};

  \draw [dashed, ->, line width=.2mm] (4.54, -1.3) -- (3.72, -1.3);
  \draw (4.1, -1.45) circle (.1cm) node {\scalebox{.5}{4}};

  \draw [dashed, ->, line width=.2mm] (1.52, -2) -- (2.48, -2);
  \draw (2, -1.85) circle (.1cm) node {\scalebox{.5}{6}};

  \draw [dashed, ->, line width=.2mm] (2.48, -2.5) -- (1.52, -2.5);
  \draw (2, -2.65) circle (.1cm) node {\scalebox{.5}{9}};

  \draw [dashed, ->, line width=.2mm] (3.72, -2) -- (7.54, -2);
  \draw (5.6, -1.85) circle (.1cm) node {\scalebox{.5}{7}};

  \draw [dashed, ->, line width=.2mm] (7.54, -2.5) -- (3.72, -2.5);
  \draw (5.6, -2.65) circle (.1cm) node {\scalebox{.5}{8}};

  \draw [dashed, <->, line width=.2mm] (nf2) -- (8.5, -3.2) -- (.9, -3.2) -- (epoxide);
  \draw (4.5, -3.05) circle (.11cm) node {\scalebox{.5}{10}};

\end{tikzpicture}
\caption{Simplified view of the UNIFY demo architecture and sequence
  of interactions.}
\label{fig:unify}
\end{figure}
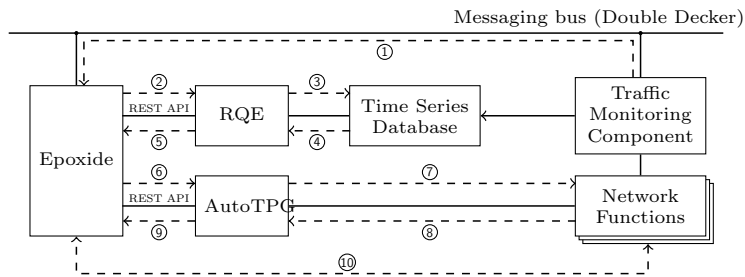

Fig.~\ref{fig:unify} highlights the main system components and the
sequence of interactions.  First, a monitoring component detects the
resource imbalance that automatically triggers the execution of the
troubleshooting process in \epoxide{} that executes a \gls{tsg}
tailored for this scenario (step~1).  In step~2, \epoxide{} asks the
Recursive Query Engine to narrow down the location of the error to a
subset of OpenFlow switches.  After querying historical measurement
data (steps~3--4), RQE returns a list of switches to \epoxide{}
(step~5).  In step~6, \epoxide{} starts and configures AutoTPG and
asks it to test candidate switches one by one.  AutoTPG tests
correctness of the switches in steps~7--8 and returns the result to
\epoxide{} in step~9.  Finally, \epoxide{} queries the flow-entries of
the erroneous switch (step~10), and presents those in tabular from to
the user to help further investigating the problem manually.

The hundred-line long \gls{tsg} of the demo contains a wrapper node
for the messaging bus, but communicates with the other tools via
\texttsg{Rest-api} nodes.  REST API calls request JSON-formatted
inputs and emit JSON-formatted outputs.  While most of the nodes
discussed so far assumed unstructured plain text inputs and outputs,
in this case we needed \texttsg{Json-filter} nodes to prepare the
outputs for post-processing, and \texttsg{Format} nodes to assemble
JSON-formatted inputs for the API calls.  This example shows that
there is not too much difference between a \gls{tsg} processing
structured JSON data and a \gls{tsg} processing unstructured data
line-by-line: similar filter nodes are necessary in both cases. The
\gls{tsg} also exemplifies the usage of a \texttsg{Tee} node and the
\texttsg{Command} node.  Similarly to its Unix counterpart,
\texttsg{Tee} copies its input text to its output link and saves the
text in a file.  \texttsg{Command} is used for running an \emph{ssh}
command to modify the configuration of a remote network function.

\epoxide{} exhibits properties that make it an ideal testing and
troubleshooting tool for Service Function Chaining.  First, the
\gls{tsg} language is an enabler of fast hypotheses testing and small
feedback cycles because it allows connecting existing special purpose
troubleshooting tools at an abstract level.  Second, the test
generation process can be further shortened by simply re-using parts
of existing \tsgFile{} definitions.  Third, complex decision logics
can be based on service-specific monitoring and troubleshooting tools
by writing simple wrapper nodes around these tools.

\section{Conclusion and Future Work}
\label{sec:conclusion}

While our modular framework proposed here is capable of flexibly
combining various troubleshooting tools for tracking down networking
issues, and the TSG concept enables the accumulation and sharing of
troubleshooting related knowledge, the current prototype
implementation should be extended in many aspects. Of course, a large
library of wrapper nodes should be added to incorporate more and more
troubleshooting tools. Besides this natural option, we outline here
some future directions, the framework could benefit from.

Although present implementation supports the addition of new node
recommenders, the currently implemented one provides only a basic
functionality.  We consider this a good basis to implement better
recommenders.  The first in line is a recommender that analyzes the
current \gls{tsg} and---by finding the most similar \gls{tsg} from a
collection---suggests new nodes based on that.  This concept could be
further improved by tagging the \glspl{tsg} as to what sort of
problems they were used for finding out and taking into account the
tags during recommendation.  Suggestions could be made more relevant
if the environment of the nodes were to be taken into consideration
as well, and node configuration arguments could be suggested also.

To support our case-based reasoning approach, we imagine a repository
of \glspl{tsg} accessible by anyone for viewing old cases and
uploading new ones.  At the same time such a repository is set up,
recommenders could be moved to the cloud as well, resulting in a more
efficient computation of suggestions on a much greater collection of
\glspl{tsg} compared to a locally computed model.

Since case-based reasoning is not the cutting edge of decision support
systems, we can use the \gls{tsg} concept as a failure detection graph
to be used as a Bayesian network, and make node suggestions based on
that.  Finding an applicable failure propagation model is key to this
approach.  We believe, we can take this concept even further by using a
Bayesian network based method and combining it with active
diagnostics---where new network tests are selected automatically depending
on current results---for \glspl{tsg} to be built with little operator
intervention or totally unsupervised.

\section*{Acknowledgment}

The research leading to these results was partly supported by Ericsson
and has received funding from the European Union Seventh Framework
Programme under grant agreement N\textsuperscript{\underline{o}}
619609.

\pagebreak

\appendix

\section{Complete \gls{tsg} Belonging to the Everyday Example}
\label{app:fulltsg}

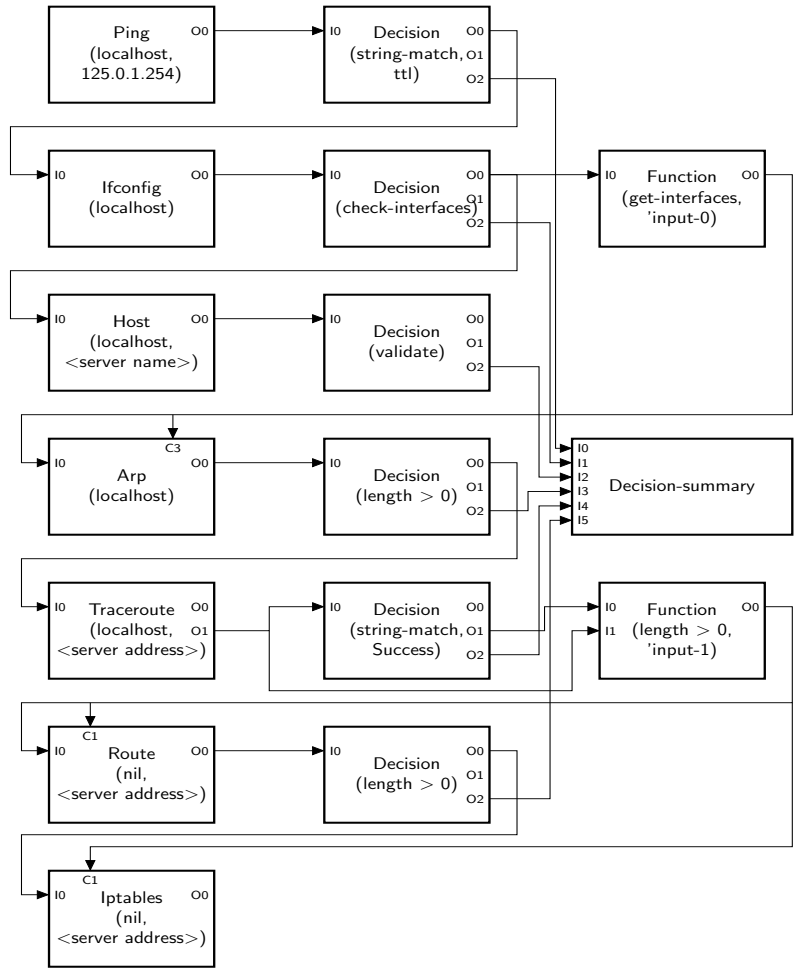
\begin{figure}[h]
\centering
\resizebox{10.5cm}{12.8cm}{%
\begin{tikzpicture}[font=\sffamily,>=triangle 45]

  \node [shape=nodea] (ping) at (1, 0) {\begin{tabular}{c} Ping \\ (localhost, \\125.0.1.254) \end{tabular}};
  \node [shape=decision] (dec1) at (6, 0) {\begin{tabular}{c} Decision \\ (string-match, \\ ttl) \end{tabular}};
  \draw [->] (ping.outputa) -- (dec1.inputa);

  \node [shape=nodeb] (ifconfig) at (1, -3) {\begin{tabular}{c} Ifconfig \\ (localhost) \end{tabular}};
  \draw [->] (dec1.outputa) -- (8, 0.5) -- (8, -1.5) -- (-1.2, -1.5) -- (-1.2, -2.5) -- (ifconfig.inputa);

  \node [shape=decision] (dec2) at (6, -3) {\begin{tabular}{c} Decision \\ (check-interfaces)\end{tabular}};
  \draw [->] (ifconfig.outputa) -- (dec2.inputa);

  \node [shape=nodeb] (function1) at (11, -3) {\begin{tabular}{c} Function \\ (get-interfaces,\\'input-0)\end{tabular}};
  \draw [->] (dec2.outputa) -- (function1.inputa);

  \node [shape=nodeb] (host) at (1, -6) {\begin{tabular}{c} Host \\ (localhost,\\\textless{}server name\textgreater{}) \end{tabular}};
  \draw [->] (dec2.outputa) -- (8, -2.5) -- (8, -4.5) -- (-1.2, -4.5) -- (-1.2, -5.5) -- (host.inputa);

  \node [shape=decision] (dec3) at (6, -6) {\begin{tabular}{c} Decision \\ (validate) \end{tabular}};
  \draw [->] (host.outputa) -- (dec3.inputa);

  \node [shape=nodec] (arp) at (1, -9) {\begin{tabular}{c} Arp \\ (localhost) \end{tabular}};
  \draw [->] (function1.outputa) -- (13, -2.5) -- (13, -7.5) -- (-1, -7.5) -- (-1, -8.5) -- (arp.inputa);
  \draw [->] (1.75, -7.5) -- (arp.confc);

  \node [shape=decision] (dec4) at (6, -9) {\begin{tabular}{c} Decision \\ (length \textgreater{} 0) \end{tabular}};
  \draw [->] (arp.outputa) -- (dec4.inputa);

  \node [shape=summary] (summary) at (11, -9) {Decision-summary};
  \draw [->] (dec1.outputc) -- (8.7, -0.5) -- (8.7, -8.2) -- (summary.inputa);
  \draw [->] (dec2.outputc) -- (8.6, -3.5) -- (8.6, -8.5) -- (summary.inputb);
  \draw [->] (dec3.outputc) -- (8.4, -6.5) -- (8.4, -8.8) -- (summary.inputc);
  \draw [->] (dec4.outputc) -- (8.2, -9.5) -- (8.2, -9.1) -- (summary.inputd);
  \draw [->] (7.5, -12.5) -- (8.4, -12.5) -- (8.4, -9.4) -- (summary.inpute);
  \draw [->] (7.5, -15.5) -- (8.6, -15.5) -- (8.6, -9.7) -- (summary.inputf);

  \node [shape=nodee] (trace) at (1, -12) {\begin{tabular}{c} Traceroute \\ (localhost, \\ \textless{}server address\textgreater{}) \end{tabular}};
  \draw [->] (dec4.outputa) -- (8, -8.5) -- (8, -10.5) -- (-1, -10.5) -- (-1, -11.5) -- (trace.inputa);

  \node [shape=decision] (dec5) at (6, -12) {\begin{tabular}{c} Decision \\ (string-match, \\ Success) \end{tabular}};
  \draw [->] (trace.outputb) -- (3.5, -12) -- (3.5, -11.5) -- (dec5.inputa);

  \node [shape=nodeg] (function2) at (11, -12) {\begin{tabular}{c} Function \\ (length \textgreater{} 0, \\'input-1) \end{tabular}};
  \draw [->] (dec5.outputb) -- (8.5, -12) -- (8.5, -11.5) -- (function2.inputa);
  \draw [->] (3.5, -12) -- (3.5, -13.25) -- (9, -13.25) -- (9, -12) -- (function2.inputb);

  \node [shape=noded] (route) at (1, -15) {\begin{tabular}{c} Route \\ (nil, \\\textless{}server address\textgreater{}) \end{tabular}};
  \draw [->] (function2.outputa) -- (13, -11.5) -- (13, -13.5) -- (-1, -13.5) -- (-1, -14.5) -- (route.inputa);
  \draw [->] (0.25, -13.5) -- (route.confa);

  \node [shape=decision] (dec6) at (6, -15) {\begin{tabular}{c} Decision \\ (length \textgreater{} 0) \end{tabular}};
  \draw [->] (route.outputa) -- (dec6.inputa);

  \node [shape=noded] (ipt) at (1, -18) {\begin{tabular}{c} Iptables \\ (nil, \\\textless{}server address\textgreater{}) \end{tabular}};
  \draw [->] (dec6.outputa) -- (8, -14.5) -- (8, -16.25) -- (-1, -16.25) -- (-1, -17.5) -- (ipt.inputa);
  \draw [->] (13, -13.5) -- (13, -16.5) -- (0.25, -16.5) -- (ipt.confa);
  
\end{tikzpicture} } % end of resizebox
\caption{\gls{tsg} for the example.}
\label{fig:everyday_example}
\end{figure}

\vfill
\pagebreak

\section{List of Nodes Currently Implemented in \epoxide{}}
\label{app:listofnodes}
\scriptsize
\renewcommand{\arraystretch}{1.3}

\begin{longtable}[h!]{>{\bfseries}p{.23\textwidth}p{10cm}} 
  \hline
    Arp: & queries the \gls{arp} cache by evoking the arp command.\\
  \hline
    Clock: & provides a signal on a timely basis.\\
  \hline
    Command: & wraps any shell commands. These are run as \emacs{} subprocesses.\\
  \hline
    Decision: & provides conditional branching.\\
  \hline
    Decision-summary: & summarizes the results of connected Decision
    nodes.\\
  \hline
    Doubledecker: & connects to a DoubleDecker broker and is able to
    receive and send messages to a partner or to a topic.\\
  \hline
    Dpids-\textless{}controller type\textgreater{}: & collects
    DPID (Datapath ID) and switch name information from an OpenFlow
    controller.\\
  \hline
    Emacs-buffer: & provides options to channel information from any
    \emacs{} buffer to \epoxide{}.\\
  \hline
    Escape: & queries NFFG topology information from Escape.\\
  \hline
    Filter: & marks incoming text that matches a given regular
    expression.\\
  \hline
    Flow-space-filter: & provides support to select only that flow
    space that is wished to be seen.\\
  \hline
    Flow-stat-\textless{}controller type\textgreater{}: &
    provides functionality to query an OpenFlow controller for flow
    statistics of a specific switch given with its DPID.\\
  \hline
    Format: & collects text from its inputs and reformats them.\\
  \hline
    Function: & wraps any \emacslisp{} function or nameless function.\\
  \hline
    Gdb: & attaches a wrapped \gls{gdb} to a running process.\\
  \hline
    Graph: & provides graph visualization support.\\
  \hline
    Host: & performs DNS lookup using the host shell-call.\\
  \hline
    Ifconfig: & wraps the ifconfig shell command.\\
  \hline
    Iperf: & wraps around the iperf command.\\
  \hline
    Json-filter: & finds the values belonging to a given key in a JSON
    expression.\\
  \hline
    Ping: & wraps the ping command.\\
  \hline
    Rest-api: & performs REST API calls.\\
  \hline
    Route: & wraps the route command.\\
  \hline
    Table-view: & displays data received on its inputs in a table
    form.\\
  \hline
    Tee: & wraps around the Unix tee command: while forwarding incoming 
    data it also saves it to a given file.\\
  \hline
    Topology-\textless{}controller type\textgreater{}: &
    queries topology information from an OpenFlow controller.\\
  \hline
    Traceroute: & wraps the traceroute process.\\
  \hline
      %     \end{tabular}
  \caption{List of implemented \epoxide{} nodes.}
\end{longtable}

\bibliographystyle{IEEEtran}
\bibliography{IEEEabrv,ms}

\end{document}